\documentclass[longauth]{aa}  

\usepackage{graphicx}
\usepackage{txfonts}
\usepackage{hyperref}
\newcommand*{\teff}{\ensuremath{T_{\mathrm{eff}}}}
\newcommand*{\mj}{\ensuremath{M_{\mathrm{J}}}}
\usepackage[utf8]{inputenc}
\usepackage{multirow}

\makeatletter
\renewcommand*\aa@pageof{, page \thepage{} of \pageref*{LastPage}}
\makeatother

\begin{document}

   \title{A scaled-up planetary system around a supernova progenitor\thanks{Based on observation
    made with European Southern Observatory (ESO) telescopes at Paranal
    Observatory in Chile, under program 1101.C0258}}


   \author{V. Squicciarini
          \inst{1,2}
          \and
          R. Gratton\inst{2}
          \and
          M. Janson\inst{3}
          \and
          E. E. Mamajek\inst{4}
          \and
          G. Chauvin\inst{5,6}
          \and
          P. Delorme\inst{5}
          \and
          M. Langlois\inst{7}
          \and
          A. Vigan\inst{8}
          \and
          S. C. Ringqvist\inst{3}
          \and
          G. Meeus\inst{9,10}
          \and
          S. Reffert\inst{11}
          \and
          M. Kenworthy\inst{12}
          \and
          M. R. Meyer\inst{13}
          \and
          M. Bonnefoy\inst{7}
          \and
          M. Bonavita\inst{14}
          \and
          D. Mesa\inst{2}
          \and
          M. Samland\inst{3}
          \and
          S. Desidera\inst{2}
          \and
          V. D'Orazi\inst{2}
          \and
          N. Engler\inst{15}
          \and
          E. Alecian\inst{5}
          \and
          A. Miglio\inst{16}
          \and
          T. Henning\inst{17}
          \and
          S. P. Quanz\inst{15}
          \and
          L. Mayer\inst{18}
          \and
          O. Flasseur\inst{19}
          \and
          G.-D. Marleau\inst{20,21,17}
          }

   \institute{Department of Physics and Astronomy ‘Galileo Galilei’, University of Padova; Vicolo dell’Osservatorio 3, I-35122 Padova, Italy\\
              \email{vito.squicciarini@inaf.it}
         \and
             INAF – Osservatorio Astronomico di Padova; Vicolo dell’Osservatorio 5, I-35122 Padova, Italy
        \and
            Institutionen för astronomi, Stockholms universitet; AlbaNova universitetscentrum, SE-106 91 Stockholm, Sweden  
        \and
            Jet Propulsion Laboratory, California Institute of Technology; 4800 Oak Grove Drive, Pasadena CA 91109, USA.
        \and
            Univ. Grenoble Alpes, CNRS, IPAG; F-38000 Grenoble, France
        \and
            Unidad Mixta Internacional Franco-Chilena de Astronomía, CNRS/INSU UMI 3386 and Departamento de Astronomía, Universidad de Chile; Casilla 36-D, Santiago, Chile
        \and
            CRAL, UMR 5574, CNRS, Université Lyon 1; 9 avenue Charles André, F-69561 Saint-Genis-Laval Cedex, France
        \and
            Aix-Marseille Université, CNRS, CNES, LAM (Laboratoire d’Astrophysique de Marseille); UMR 7326, F-13388 Marseille, France.
        \and
            Departamento de Física Teórica, Universidad Autónoma de Madrid; S-28049 Madrid, Spain
        \and
            Centro de Investigación Avanzada en Física Fundamental (CIAFF), Facultad de Ciencias, UAM; S-28049 Madrid, Spain
        \and
            Landessternwarte, Zentrum für Astronomie der Universität Heidelberg; Königstuhl 12, D-69117 Heidelberg, Germany
        \and
            Leiden Observatory, Leiden University; Postbus 9513, NL-2300 RA Leiden, The Netherlands
        \and
            Department of Astronomy, University of Michigan; 1085 S. University Ave, Ann Arbor MI 48109, USA
        \and
            School of Physical Sciences, Faculty of Science, Technology, Engineering and Mathematics, The Open University; Walton Hall, Milton Keynes, MK7 6AA, UK
        \and
            ETH Zürich, Institute for Particle Physics and Astrophysics; Wolfgang-Pauli-Strasse 27, CH-8093 Zürich, Switzerland
        \and
            School of Physics and Astronomy, University of Birmingham; Edgbaston, B15 2TT, UK
        \and
            Max-Planck-Institut für Astronomie, Königstuhl 17, D-69117 Heidelberg, Germany
        \and
            Center for Theoretical Physics and Cosmology, Institute for Computational Science, Universität Zürich; Winterthurerstrasse 190, CH-8056 Zürich, Switzerland
        \and
            LESIA, Observatoire de Paris, Université PSL, CNRS, Sorbonne Université, Univ. Paris Diderot; Sorbonne Paris Cité, 5 Place Jules Janssen, F-92195 Meudon, France
        \and
            Institut für Astronomie und Astrophysik, Universität Tübingen; Auf der Morgenstelle 10, D-72076 Tübingen, Germany
        \and
            Physikalisches Institut, Universität Bern; Gesellschaftsstr. 6, CH-3012 Bern, Switzerland
             }

   \date{Received --; accepted --}

  \abstract
   {
   Virtually all known exoplanets reside around stars with $M<2.3~M_\odot$ either due to the rapid evaporation of the protostellar disks or to selection effects impeding detections around more massive stellar hosts.}
   {To clarify if this dearth of planets is real or a selection effect, we launched the planet-hunting B-star Exoplanet Abundance STudy (BEAST) survey targeting B stars ($M>2.4~M_\odot$) in the young (5-20 Myr) Scorpius-Centaurus association by means of the high-contrast spectro-imager SPHERE at the Very Large Telescope.}
   {In this paper we present the analysis of high-contrast images of the massive ($M \sim 9~M_\odot$) star $\mu^2$ Sco obtained within BEAST. We carefully examined the properties of this star, combining data from Gaia and from the literature, and used state-of-the-art algorithms for the reduction and analysis of our observations.}
   {Based on kinematic information, we found that $\mu^2$ Sco is a member of a small group which we label Eastern Lower Scorpius within the Scorpius-Centaurus association. We were thus able to constrain its distance, refining in turn the precision on stellar parameters. Around this star we identify a robustly detected substellar companion ($14.4\pm 0.8$~\mj) at a projected separation of $290\pm 10$~au, and a probable second similar object ($18.5\pm 1.5$~\mj) at $21\pm 1$~au. The planet-to-star mass ratios of these objects are similar to that of Jupiter to the Sun, and the flux they receive from the star is similar to those of Jupiter and Mercury, respectively. }
   {The robust and the probable companions of $\mu^2$ Sco are naturally added to the giant $10.9~\mj$ planet recently discovered by BEAST around the binary b Cen system. While these objects are slightly more massive than the deuterium burning limit, their properties are similar to those of giant planets around less massive stars and they are better reproduced by assuming that they formed under a planet-like, rather than a star-like scenario. Irrespective of the (needed) confirmation of the inner companion, $\mu^2$ Sco is the first star that would end its life as a supernova that hosts such a system. The tentative high frequency of BEAST discoveries is unexpected, and it shows that systems with giant planets or small-mass brown dwarfs can form around B stars. When putting this finding in the context of core accretion and gravitational instability formation scenarios, we conclude that the current modeling of both mechanisms is not able to produce this kind of companion. The completion of BEAST will pave the way for the first time to an extension of these models to intermediate and massive stars.}

   \keywords{planetary systems --
                Stars: early-type --
                Stars: individual: \object{mu2 Sco} --
                Stars: individual: \object{b Cen} --
                Techniques: high angular resolution --
               }

   \maketitle
%

\section{Introduction}
\label{sec:introduction}

Prior to the discovery of exoplanets, the Solar System appeared to be the inevitable outcome of a smooth process of planetary formation: a process going from the primordial protoplanetary disk to the present ordered sequence of planets with nearly circular orbits and a tendency for a decrease in density with distance, such that the rocky planets are followed by giant planets as a consequence of the initial temperature gradient in the disk itself \citep{1985prpl.conf.1100H}. However, the detection of the first exoplanet around a solar-type star \citep{1995Natur.378..355M} thoroughly challenged this view. With a mass of $0.5 \mj$, 51 Pegasi b lies so close to its star that it completes one orbit in 4.2 days. This planet became the archetype of a whole new family of planets, the Hot Jupiters. In the following 27 years, the discovery of exoplanets with properties markedly different from the ones in our system -- that is to say extremely eccentric orbits, orbits perpendicular to the equatorial plane of their star, or evaporating atmospheres -- has profoundly changed our view of planet formation and evolution \citep{2015ARA&A..53..409W}. Indeed, the current view is that of a complex process involving mechanisms such as mutual resonances, orbital migration, and scattering \citep[see, e.g.,][]{2017A&A...602A.107B}, which emphasizes in turn the need for larger surveys to investigate the diversity we see systematically.
    
Despite advancing at a steady pace, laying the foundations of (exoplanet) demographics, a comprehensive and unbiased census of the planetary population is still missing, since every detection technique is more sensitive to certain regions of the parameter space\footnote{Which comprises, but is not limited to, planetary (mass, radius), stellar (mass, radius, age, activity), and orbital (orbital distance, period, eccentricity, inclination) properties.} \citep{2021exbi.book....2G}. Most previous and ongoing surveys have focused on stars as massive as (or less massive than) the Sun\footnote{More massive stars than the Sun have mostly been targeted in later evolutionary phases, once they have left the main sequence to become giant stars.}, and about 90\% of the 5000 known exoplanets lie closer to their host stars than the Earth is to the Sun. For a few years, this strong observational bias has started being alleviated by direct imaging which, in contrast to the prevailing radial velocity and transit methods, is more sensitive to thermal emission from giant planets orbiting far from their stars. Since at young ages ($\lesssim 1$ Gyr) giant planets are mainly heated by gravitational contraction rather than by stellar radiation, they fade in brightness over time: the best targets are therefore young and nearby stars.
    
The few previous direct imaging and radial velocity surveys targeting more massive stars than the Sun \citep[e.g.,][]{2010PASP..122..905J,2015A&A...574A.116R,2022AJ....163...80W} have shown that gas giant planets are more frequent around more massive stars, likely due to the increased dust and gas reservoir in protoplanetary disks \citep{2012A&A...541A..97M}. According to radial velocity studies, however, the occurrence frequency has a turnover at about 2~$M_\odot$ and goes down to zero at $M >3 M_\odot$ \citep{2015A&A...574A.116R}, at least for those orbiting within a few au of their star. This is in line with theoretical expectations from the most popular scenario for planet formation, the core accretion model \citep[CA, ][]{1996Icar..124...62P}. This mechanism, which is thought to be responsible for the formation of the giant planets in the Solar System, requires the growth of solid planetary cores from dust, a slow process requiring a few million years. Therefore, due to a more rapid dispersal of the protoplanetary disk around heavier stars, giant planet formation should be increasingly hampered—and eventually halted around more massive stars \citep{2012A&A...541A..97M}.
    
Radial velocity surveys, though, are strongly biased toward close-in planets. Unlike CA, gravitational instability \citep[GI, ][]{2003ApJ...599..577B} may well form massive planets and substellar objects at a wide separation on a timescale shorter ($\sim 10^4$ yr) than that the lifetime of the disk. In addition, the ratio between the mass of the disk and that of the star is known to increase with stellar mass \citep{2016ApJ...831..125P}, favoring the onset of GI \citep{2020MNRAS.492.5041C}. If GI is a viable formation pathway for giant planets, the dearth of planetary-like companions around massive stars would be due to selection effects inherent to radial velocity searches.

On the other hand, as already mentioned, massive planets at large separations can be detected through direct imaging provided that they are young enough. Past high-contrast imaging surveys \citep[e.g.,][]{2013A&A...553A..60R,2016PASP..128j2001B,2017sf2a.conf..331C,2019ESS.....410002N,2021A&A...651A..72V} have focused on stars of spectral type A or later ($M \lesssim 2.4 M_\odot$); to fill in the gap, in 2018 we started the B-star Exoplanet Abundance STudy (BEAST) survey \citep{2021A&A...646A.164J}, exploiting the capabilities of the Spectro-Polarimetric High-contrast Exoplanet Research (SPHERE) instrument \citep{2019A&A...631A.155B} at the Very Large Telescope (VLT): its scientific goal is to look for exoplanets around a sample of 85 B stars in the young (5–20 million years) Scorpius-Centaurus OB association (Sco-Cen), the nearest region of ongoing stellar formation. While the survey is still in progress, we recently reported the discovery of a $10.9 \pm 1.6 \mj$ planet around the stellar binary b Cen, whose total mass reaches 6-10~M$_\odot$ \citep{2021Natur.600..231J}.
    
Here we present a second planetary system surrounding the massive star $\mu^2$ Scorpii ($\mu^2$ Sco), composed of a robustly-detected $14.4 \pm 0.8~ \mj$ companion and a probable closer $18.5 \pm 1.5~ \mj$ companion, and discuss its implication for the current scenarios of planetary formation. We present an extensive discussion of stellar properties in Section~\ref{sec:star}. Section~\ref{sec:observations} details the observations and Section~\ref{sec:data_reduction} discusses the data reduction methods. In Section~\ref{sec:cc_candidates} we identify the companion candidates, treating with special care the innermost one. In Section~\ref{sec:cc_analysis} we discuss the possibility that the two apparently comoving candidates are not physically bound to the star. Aftering ruling out this possibility, we derive their properties and constrain their orbits. In Section~\ref{sec:discussion} we discuss the properties of these low-mass ratio companions in a more general context and examine the implications of their detection on the origin of this class of companions around B stars.


\section{The star}
\label{sec:star}

$\mu^2$ Sco (also known as HR 6252, HD 151985, HIP 82545, Pipirima\footnote{Pipirima is the Tahitian name for the pair $\mu^2$ and $\mu^1$ Sco, referring to mythological twin siblings. The IAU Working Group on Star Names (WGSN) adopted Pipirima as the proper name for $\mu^2$ Sco in 2017 (https://www.iau.org/public/themes/naming\_stars/).}) is a naked-eye star belonging to the young Scorpius-Centaurus association \citep{1999AJ....117..354D}. The main astrometric, kinematic, and photometric properties of the star are reported in Table~\ref{table:mu2sco_data}.

Given that several stellar parameters (notably distance, age and mass) are of utmost importance for the characterization of directly-imaged exoplanets, our primary goal was to reduce the large uncertainty on distance that had historically limited a self-consistent physical analysis of $\mu^2$ Sco. After exploiting for the first time kinematic information to indirectly constrain its distance, we combined this new information with data from the literature to perform a Monte Carlo analysis that determines posterior distributions for stellar mass, age, radius and effective temperature. A discussion on the adopted priors is provided in Section~\ref{sec:distance}-\ref{sec:surface_gravity}, and the final derived parameters are reported in Section~\ref{sec:final_param}. Complete details on the derivation are provided in Appendix~\ref{sec:optimization}.

\begin{table}
\caption{Main astrometric, kinematic and photometric properties of $\mu^2$ Sco collected from the literature. References: (1): \citet{2021A&A...649A...1G}; (2): \citet{2019A&A...623A..72K}; (3): \citet{2006AstL...32..759G}; (4): \citet{1997A&A...319..811B}; (5) \citet{2006yCat.2168....0M}; (6): \citet{2003yCat.2246....0C}; (7) \citet{1969ApJ...157..313H}.}
\label{table:mu2sco_data}
\centering
\begin{tabular}{ccc}
\hline\hline
Name & Value & Reference \\
\hline        
$\alpha$ ($^\circ$, J2016.0) & 253.083869820(64) & (1) \\
$\delta$ ($^\circ$, J2016.0) & -38.017636551(38) & (1) \\
$\mu_\alpha^*$ (mas yr$^{-1}$) & $-11.772 \pm 0.022$ & (2) \\
$\mu_\delta$ (mas yr$^{-1}$) & $-23.105 \pm 0.021$ & (2) \\
RV (km s$^{-1}$) & $1.3 \pm 0.8$ & (3) \\
$v \sin{i}$ (km s$^{-1}$) & $52 \pm 5$ & (4) \\
V (mag) & $3.565 \pm 0.005$ & (5) \\
G (mag) & $3.543 \pm 0.003$ & (1) \\
J (mag) & $4.15 \pm 0.28$ & (6) \\
H (mag) & $4.159 \pm 0.25$ & (6) \\
K (mag) & $4.292 \pm 0.31$ & (6) \\
spectral type & B2IV & (7) \\
\hline 
\end{tabular}
\end{table}

\subsection{Distance and membership to the Eastern Lower Scorpius group}
\label{sec:distance}

The Scorpius-Centaurus association, the nearest extended region with ongoing stellar formation to the Sun \citep{1999AJ....117..354D}, is classically divided into three subregions: Upper Scorpius (US), Upper Centaurus-Lupus (UCL) and Lower Centaurus-Crux (LCC) \citep{1999AJ....117..354D}. Each region in turn has a complex morphology with a high degree of substructuring. $\mu^2$ Sco, in particular, appears to lie within a small clump of stars (a region centered on galactic coordinates $(l, b) = (343.1^\circ, 4.7^\circ)$ with a radius of about $2^\circ$ that, while classically assigned to UCL \citep{1999AJ....117..354D}, has recently started being recognized in its own right as an independent entity \citep{2018A&A...614A..81R}. We refer to this group as Lower Scorpius (LS).

Starting from the Gaia DR2-based catalog of bona fide Sco-Cen sources by \citet{2019A&A...623A.112D}, we create a census of LS stars (575 members) through the criteria shown in Table~\ref{tab:criteria}. We also crop from the same catalog an UCL sample, to be used for comparison purposes, using the classical boundaries by \citet{1999AJ....117..354D} and excluding the sources already assigned to LS.

\begin{figure}
\centering
\includegraphics[width=\hsize]{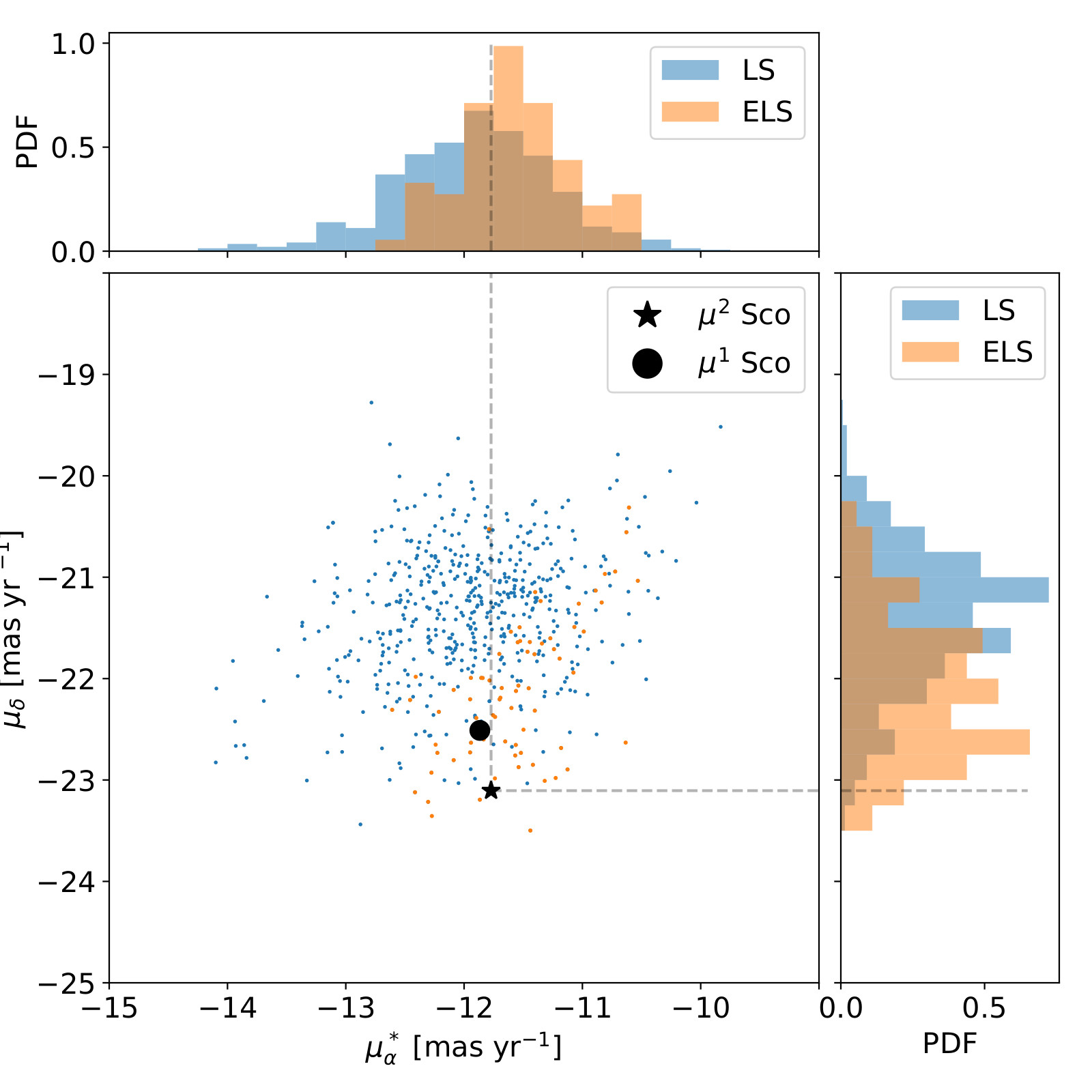}
   \caption{Proper motion components of LS (blue) and ELS (orange) members. Both $\mu^1$ Sco and $\mu^2$ Sco are consistent with membership to ELS.}
\label{fig:sample_pm}
\end{figure}

Inspection of the Gaia EDR3 catalog \citep{2021A&A...649A...1G} reveals an inconsistency of the star's proper motions with those of LS and a parallax that, although consistent with LS, suffers from a large uncertainty. The values of Gaia EDR3, however, are also inconsistent with those from Gaia DR2 and Hipparcos \citep{1997A&A...323L..49P} (Table~\ref{table:astrom_data}). The inaccuracy of the astrometric solution in Gaia for such a bright star (which saturates the detector) is expected without invoking the presence of an unresolved stellar companion (Appendix~\ref{sec:binarity}). Adopting the robust long-term proper motion by \citet{2019A&A...623A..72K}, computed as the difference between the astrometry of Hipparcos (J1991.25) and Gaia DR2 (J2015.5), the star is fully consistent with membership to LS (Figure~\ref{fig:sample_pm}). 

\begin{figure*}
\centering
\includegraphics[width=\hsize]{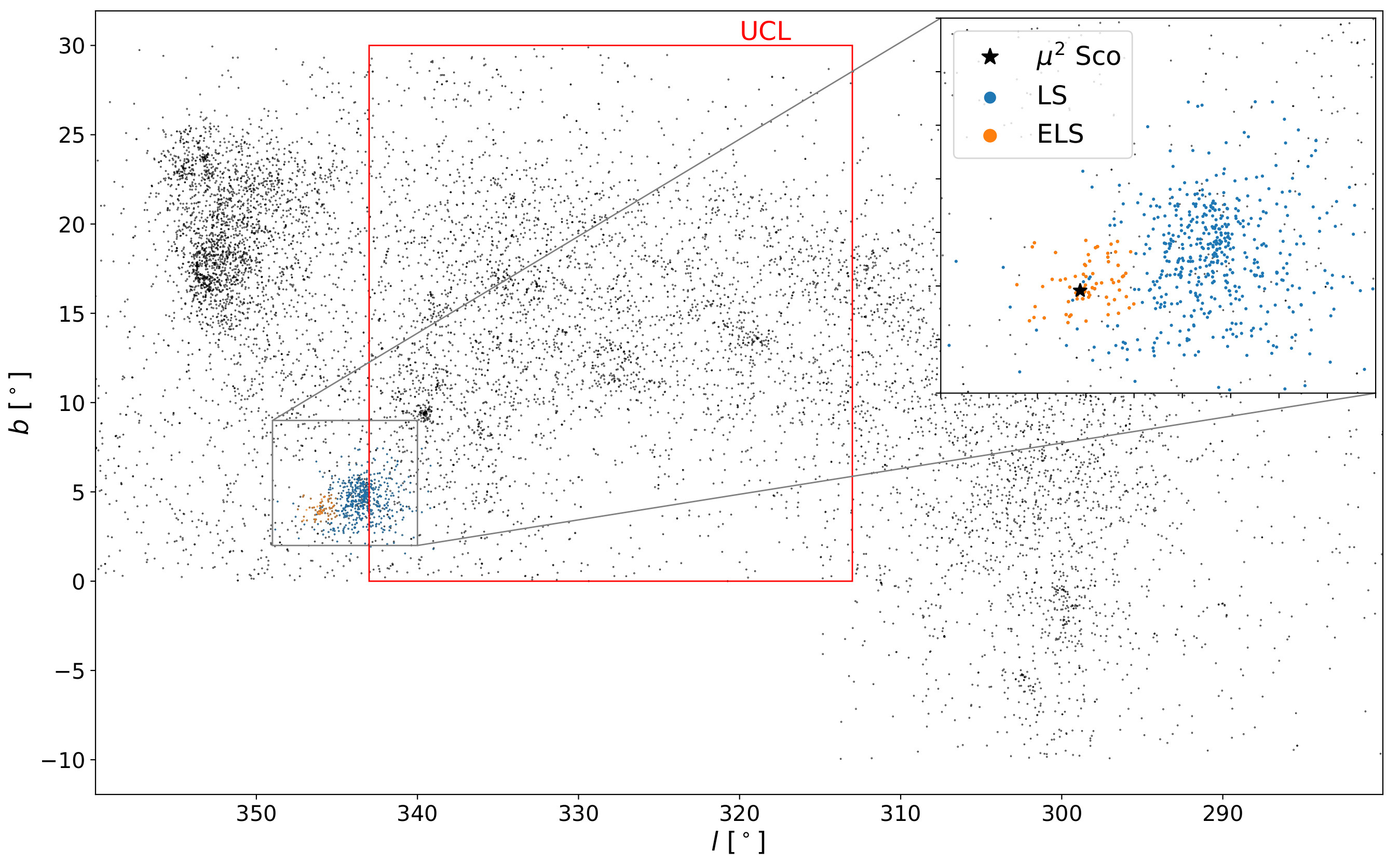}
   \caption{Sky coordinates of bona fide Sco-Cen members from Damiani et al. (2019), shown as black dots. Classical boundaries for UCL are indicated by the red box. A zoom of LS (blue) and ELS (orange) region is drawn in the upper-right corner; $\mu^2$ Sco is indicated by the black star.}
\label{fig:sample_radec}
\end{figure*}

\begin{figure}
\centering
\includegraphics[width=\hsize]{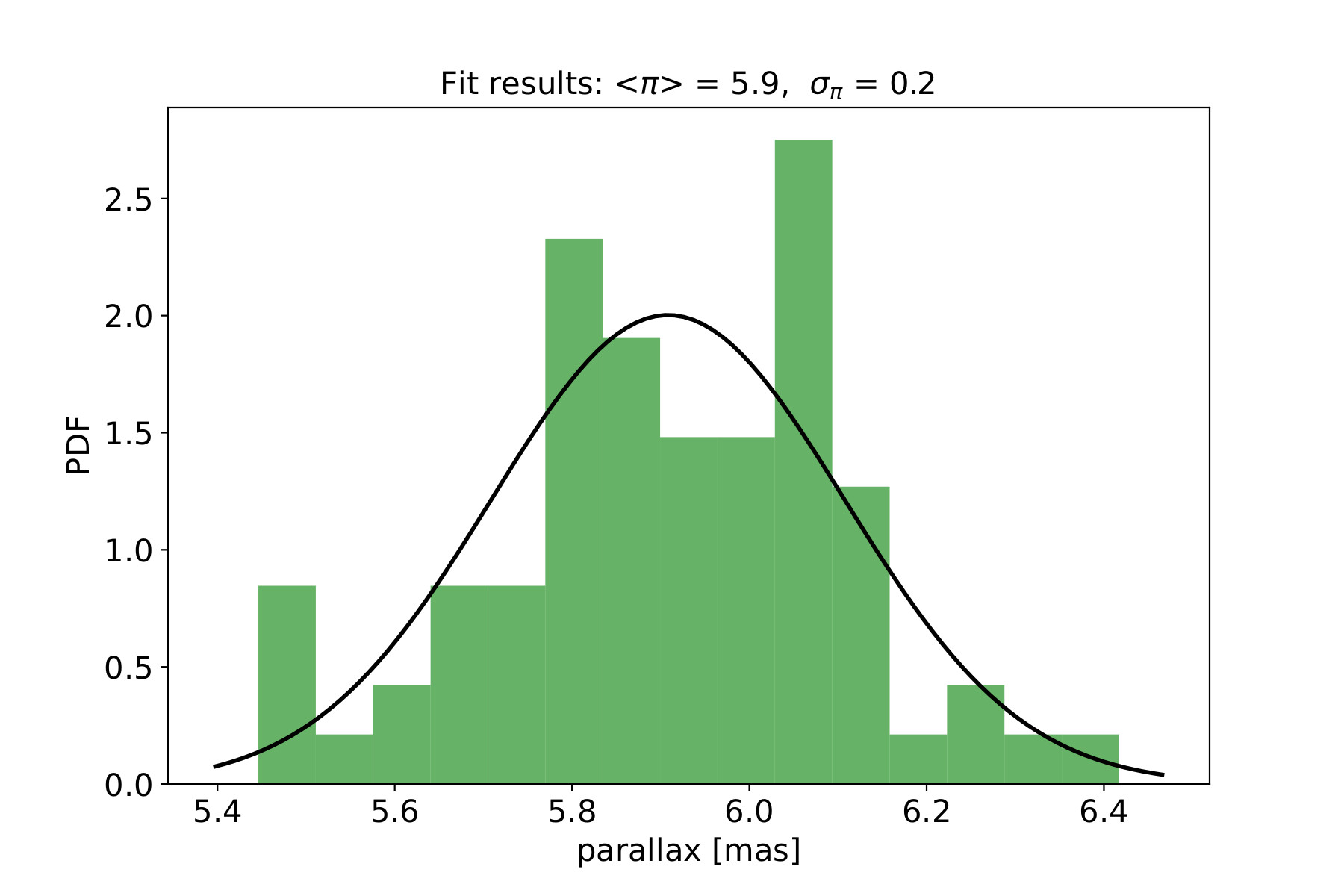}
   \caption{Parallax distribution of ELS stars. We adopt $\pi=5.9\pm0.2$ mas as parallax prior for $\mu^2$ Sco.}
\label{fig:sample_par}
\end{figure}

We notice that our target (together with its sibling $\mu^1$ Sco) actually lies in a peripheral area of LS, $\sim 2^\circ$ eastward of its core: using the more sensitive Gaia EDR3 to expand the census of LS to fainter stars, we constructed a catalog of members of this small clump that includes 73 stars, that we call Eastern Lower Scorpius (ELS; see Table~\ref{tab:criteria}, Figure~\ref{fig:sample_radec}). ELS appears somewhat closer to the Sun than the whole LS: we adopt its parallax distribution (modeled as a normal, $\pi=5.9\pm 0.2$ mas) as our parallax prior (Figure~\ref{fig:sample_par}).

\begin{table*}
  \centering
   \caption{Defining criteria for the UCL, LS and ELS samples used throughout this paper. The criteria should be interpreted as an addition to the cuts already applied in the definition of the initial catalog. Equatorial coordinates and proper motions refer to the ICRS at epoch J2016.0.}
   \begin{tabular}{lccc} \hline \hline
   Criterion & UCL & LS & ELS \\
            \hline
  \multirow{2}{*}{initial catalog} & {\citet{2019A&A...623A.112D}}, & {\citet{2019A&A...623A.112D}}, & \multirow{2}{*}{Gaia EDR3} \\
  & bona fide sources & bona fide sources & \\
  galactic longitude $(^\circ)$ & $313<l<343$ & $339<l<350$ & $345<l<347.5$ \\
  galactic latitude $(^\circ)$ & \textemdash & $1.5<b<7.7$ & $3.3<b<5.0$ \\
  right ascension $(^\circ)$ & \textemdash & \textemdash & $251<\alpha<255$ \\
  declination $(^\circ)$ & \textemdash & \textemdash & $-40<\alpha<-36$ \\
  parallax (mas) & \textemdash & $5.3<\pi<6.2$ & $5.3<\pi<6.5$ \\
  proper motion along $\alpha$ (mas yr$^{-1}$) & \textemdash & \textemdash & $-18.7<\mu_\alpha^*<-10.2$ \\
  proper motion along $\delta$ (mas yr$^{-1}$) & \textemdash & \textemdash & $-25.0<\mu_\delta<-16.1$ \\
  velocity along $\alpha$ (km s$^{-1}$) & \textemdash & $-11.2<v_\alpha<-8.5$ & $-10<v_\alpha<-7$ \\
  velocity along $\delta$ (km s$^{-1}$) & \textemdash & $-19.3<v_\delta<-16.5$ & $-21<v_\alpha<-16$ \\
  \hline
  no. of sources & 3842 & 575 & 73 \\
  \hline\hline    
   \end{tabular}
   \label{tab:criteria}
\end{table*}

\subsection{Mass and radius}
\label{sec:mass}

The mass estimate in the first BEAST paper \citep{2021A&A...646A.164J} was based on an underestimated distance from Gaia DR2: we therefore expect a higher mass than previously assumed. We conservatively start from a uniform prior distribution, $ M \in [5,12] M_\odot$. Likewise, we do not favor a particular initial value for the radius: $R \in [3.5,10.0]~R_\odot$. For a B2V star on the main sequence we expect $R \sim 4$~R$_\odot$ \citep{2013ApJS..208....9P}, but the classification of the star as a B2IV \citep{1969ApJ...157..313H} argues for a larger radius.

\subsection{Age}
\label{sec:age}

As already mentioned, the derivation of a reliable stellar age is crucial in the context of the characterization of directly-imaged planets, since it is needed to turn the observed fluxes into mass estimates. Direct isochronal age determinations for B stars are especially difficult: their luminosity has only a weak dependence on age because they are already on the main sequence after $\sim 10^5$ yr, and deviations from theoretical expectations might arise due to unresolved companions, strong magnetic fields and/or rapid rotation\citep{2016EAS....80..115B}.
    
An indirect dating technique, based on isochronal analysis of comoving stars rather than the target itself, can be envisioned, in a similar way as in \citet{2021MNRAS.507.1381S}: using the MADYS tool (Squicciarini et al., in preparation), we collected photometry from Gaia DR2/EDR3 and 2MASS \citep{2003yCat.2246....0C} for both LS and ELS and compared it to pre-MS isochrones to simultaneously derive individual mass and age estimates for their members \citep{2021MNRAS.507.1381S}. To avoid known systematic uncertainties on absolute ages with this method \citep[see discussion in ][]{2021MNRAS.507.1381S}, we decided to employ the whole UCL ($\sim 4000$ stars) as a control sample, comparing the derived age distributions with each other.
    
The age distribution of ELS stars, derived by comparison with the BHAC15 isochrones \citep{2015A&A...577A..42B} of solar metallicity, is consistent with that of the whole UCL according to a Kolmogorov-Smirnov test ($\alpha=0.05$)\footnote{The higher ($\sim 5$ and $\sim 4$ times, respectively) density of LS and ELS with respect to the coeval UCL is likely due to significant self-gravity, slowing down the expansion caused by random turbulent motion within the natal molecular cloud. Indeed, the dispersion of proper motions of ELS (0.71 mas yr$^{-1}$ $\approx 0.55$ km s$^{-1}$ is lower than the escape velocity of $\sim 0.72$ km s$^{-1}$ obtained by summing the masses of all observed members ($\sim 50~M_\odot$) and considering its projected size ($\rho \sim 17' \sim 0.85$ pc). The lifetime of a small cluster such as ELS against tidal disruption is about 35 Myr \citep{2005A&A...441..117L}, about twice its estimated age.}.
    
In the absence of tighter constraints, we adopt the age of UCL ($t = 16 \pm 7$ Myr: \citealt{2016MNRAS.461..794P}) as our age prior.

We highlight that an independent age estimate could come in the next years from asteroseismology: based on its light curve from TESS \citep{2015JATIS...1a4003R}, the star shows evidence for $\beta$ Cephei type variability and in particular for that of a slowly pulsating B-star (SPB) $\beta$ Cep hybrid. A quantitative analysis of the observed wealth of pulsational modes -- especially if complemented by new observations in different photometric bands -- might constitute the subject of a future work aimed at estimating the core hydrogen fraction, closely related to stellar age.

\subsection{Reddening}
\label{sec:reddening}
    
To estimate the reddening $E(B-V)$ in the direction of $\mu^2$ Sco in a robust way, we considered six different determinations:
\begin{itemize}
    \item starting from $UBV$ photometry from \citet{2006yCat.2168....0M} and $uvby\beta$ photometry from \citet{1997yCat.2215....0H}, we employ the Q-method of \citet{1953ApJ...117..313J} to deredden OB stars with the modern calibration of \citet{2013ApJS..208....9P}: $E(B-V) = 0.022$ mag;
    \item dereddening the $uvby\beta$ photometry through the calibration of \citet{1991A&A...251..106C}: $E(B-V)=0.019$ mag;
    \item dereddening the $uvby\beta$ photometry with the older, independent calibration of \citet{1983MNRAS.205.1215S}: $E(B-V)=0.020$ mag;
    \item from Lyman $\alpha$ observations, \citet{1974ApJ...191..659S} estimate the interstellar column density of HI toward $\mu^2$ Sco to be $N(H) = 2.5 \cdot 10^{20} \text{ cm}^{-2}$. Adopting the recent relation between interstellar reddening and hydrogen column densities by \citet{2017ApJ...846...38L}, valid in the low column-density regime as in our case: $E(B-V)=0.031$ mag;
    \item starting from the absorption EW(D1) of interstellar neutral sodium (NaI) D1 line toward $\mu^2$ Sco by \citet{1978ApJ...222..491H}, and using the reddening vs EW(D1) correlation by \citet{2012MNRAS.426.1465P}: $E(B-V)=0.019$ mag;
    \item integrating the STILISM 3D reddening map \citep{2018A&A...616A.132L} along the line of sight up to a distance $d=165$ pc: $E(B-V)=0.025\pm 0.024$ mag. 
\end{itemize}
By averaging these estimates, we obtain $E(B-V)=0.022 \pm 0.006$ mag. Adopting a total-to-selective extinction ratio $R_V = A_V/E(B-V) = 3.05$, appropriate for early B-type stars \citep{2004AJ....128.2144M}, we estimate $A(V) = 0.068 \pm 0.015$ mag and $A(K) = 0.0062 \pm 0.0014$. Being negligible with respect to photometric errors, extinction values will be from this moment on treated as constants.
    
\subsection{Effective temperature}
\label{sec:teff}

Concerning the stellar effective temperature, $\mu^2$ Sco -- persistently classified as a B2IV star \citep{1969ApJ...157..313H} -- has a nearly identical combination of colors and reddening ($B-V = -0.214$, $U-B = -0.844$, $E(B-V) = 0.022$) to the Morgan-Keenan B2IV standard star $\delta$ Cet ($B-V = -0.219$, $U-B = -0.850$, $E(B-V) = 0.018$) which has a median $\teff$ in the recent literature of $\teff \simeq 21600$ K \citep{2021MNRAS.504.3730C}. Hence we do not expect the effective temperature of $\mu^2$ Sco to differ too much from that of $\delta$ Cet.

\begin{table}
\caption{New effective temperature estimates for $\mu^2$ Sco. Details on the derivation of each estimate are provided in the text.}
\label{table:teff_estimates}
\centering
\begin{tabular}{c c}
\hline\hline
\# & $\teff$ (K) \\
\hline        
   1 & 20913 \\  
   2 & 21083 \\
   3 & 21655 \\
   4 & 21989 \\
   5 & 22063 \\ 
   6 & $22700\pm270$ \\ 
   7 & 25978 \\ 
\hline 
   adopted & $21900\pm1000$ \\ 
\hline          
\end{tabular}
\end{table}

In Table~\ref{table:teff_estimates} we list several new $\teff$ estimates for $\mu^2$ Sco based on photometric data from the literature.
The estimates are derived in the following way:
\begin{enumerate}
    \item photometry from \citet{1997yCat.2215....0H}, dereddened through \citet{1991A&A...251..106C} using \citet{1994MNRAS.268..119B} $\teff$ calibration;
    \item  photometry from \citet{1997yCat.2215....0H}, $\teff/H_\beta$ relation by \citet{1984MNRAS.211..973B};
    \item  employing the $(U-B)_o$ vs $\teff$ trend based on B2IV standard stars \citet{2013ApJS..208....9P};
    \item  photometry from \citet{1997yCat.2215....0H}, dereddened through \citet{1991A&A...251..106C} using \citet{1993A&A...268..653N} $\teff$ calibration;
    \item  photometry from \citet{1997yCat.2215....0H}, using [c1] index adopting \citet{2013A&A...550A..26N} $\teff$ scale;
    \item comparing IUE spectrophotometry\footnote{Taken from \href{https://archive.stsci.edu/iue/}{https://archive.stsci.edu/iue/}.} in the wavelength range [110-195] nm with the grid of model atmospheres by \citet{2003IAUS..210P.A20C};
    \item  photometry from \citet{1997yCat.2215....0H}, using [u-b] index adopting \citet{1989A&A...216...44D} $\teff$ scale.
\end{enumerate}
We adopted the averaged value ($\teff =21900\pm1000$ K) as our $\teff$ prior.

\subsection{Surface gravity}
\label{sec:surface_gravity}
The stellar surface gravity was not used as a free parameter, but rather as one of the observational constraints for the optimization tests. Based on literature estimates (Table~\ref{table:logg_estimates}), we adopted a value of $\log{g} \in [3.6,4.0]$.

\begin{table}
\caption{Literature surface gravity estimates for $\mu^2$ Sco. References for each estimate are provided in the first column. References: (1): \citet{1989A&A...216...44D}; (2): \citet{1990AJ....100.1994W}; (3): \citet{1992ApJS...78..205G}; (4): \citet{2013A&A...550A..26N}.}
\label{table:logg_estimates}
\centering
\begin{tabular}{c c}
\hline\hline
Source & $\log{g}$ \\
\hline        
   (1) & $3.9\pm0.2$ \\  
   (2) & 3.9 \\
   (3) & $3.6\pm0.2$ \\
   (4) & 3.916 \\
\hline 
   adopted & $3.8\pm0.2$ \\ 
\hline          
\end{tabular}
\end{table}

\subsection{Final stellar parameters}
\label{sec:final_param}

While we refer to Appendix~\ref{sec:optimization} for extensive details on the optimization analysis, we list in Table~\ref{table:fitted_param} the final stellar parameters obtained for $\mu^2$ Sco. 

\begin{table}
\caption{Stellar parameters of $\mu^2$ Sco derived in this work.}
\label{table:fitted_param}
\centering
\begin{tabular}{cc}
\hline\hline
Name & Value \\
\hline 
    parallax (mas) & $5.9 \pm 0.2$ \\
    $\teff$ (K) & $21700 \pm 900$ \\
    $E(B-V)$ (mag) & $0.022 \pm 0.006$ \\
    $A(V)$ (mag) & $0.068 \pm 0.015$ \\
    age (Myr) & $20 \pm 4$ \\
    mass ($M_\odot$) & $9.1 \pm 0.3$ \\
    radius ($R_\odot$) & $5.6 \pm 0.2$ \\
\hline          
\end{tabular}
\end{table}

With a spectral type of B2IV, corresponding to a best-fit effective temperature of $21700\pm900$ K, $\mu^2$ Sco is one of the brightest stars in the BEAST sample \citep{2021A&A...646A.164J}. Indeed, its fitted mass of $M=9.1\pm 0.3~M_\odot$ qualifies the star, from a stellar evolution standpoint, as a "massive star" ($M>8 M_\odot$): a star that could explode as a supernova in the next 10–20 million years \citep{2017hsn..book..483N}, in particular as an electron-capture supernova \citep{1984ApJ...277..791N}. As a $20 \pm 4$ Myr star, $\mu^2$ Sco is currently evolving off the main sequence, consistently with its spectral classification.

We note that in the specific case of $\mu^2$ Sco, the use of model isochrones -properly constrained to reduce degeneracies- should be considered reliable, given that the star does not show any evidence for unresolved companions (see Appendix~\ref{sec:binarity}), strong magnetic field or rapid rotation (see Appendix~\ref{sec:magnetic_field}).

\section{Observations}
\label{sec:observations}
    
High-contrast near-infrared images and spectra of $\mu^2$ Sco were acquired by means of SPHERE's dual-band imager \citep[IRDIS,][]{2008SPIE.7014E..3LD} and integral field spectrograph \citep[IFS,][]{2008SPIE.7014E..3EC} at the ESO Very Large Telescope \citep{2019A&A...631A.155B}. The first epoch was obtained on April 24th, 2018, the second one on June 4th, 2021 (Table~\ref{table:obs_log}). Both observations were carried out in the so-called IRDIFS-EXT pupil-tracking mode and are diffraction limited. In this mode, IRDIS was used in the dual-band mode ($\lambda_{K_1}=2.1025\pm 0.1020~\mu$m, $\lambda_{K_2}=2.2550\pm~0.1090~\mu$m; see \citealt{2010MNRAS.407...71V}) over a field of view (FoV) of $11"\times11"$, while IFS collected images -- which can be combined to build low-resolution spectra ($R=30$) -- in the $\lambda \in [0.96-1.64]~\mu$m wavelength range for each position over a field of about $1.7" \times 1.7"$, Nyquist sampled at the shortest wavelength. Once extracted with the DRH data reduction pipeline \citep{2008ASPC..394..581P}, each spectrum is 39 pixel long. The APLC2 coronagraph \citep{2011ExA....30...39C} was used, masking the star out to a radius of 92.5 mas. As for all observations in the BEAST survey, calibration observations were taken together with science exposures: they included a flux calibration, allowing normalization to the peak of the star image, obtained by offsetting the star position off the coronagraphic mask with a suitable neutral density filter to avoid saturation of the image; a center calibration, obtained by imprinting a sinusoidal pattern to the deformable mirror, providing satellite images of the star; and empty sky exposures centered a few arcsec from the star position. Details of the observations are provided in Table~\ref{table:obs_log}.

\begin{table*}
\caption{Details of the observations of $\mu^2$ Sco. DIT is the integration time, $N_{exp}$\ the number of frames after selection (out of 192 frames), Rot is the total FoV rotation. The seeing and the Strehl ratio are averages over the duration of the exposures.}
\label{table:obs_log}
\centering
\begin{tabular}{ccccccccccc}
\hline\hline
UT Date & Instr. & Filter & DIT & $N_{exp}$ & Rot & Seeing & Strehl & Airmass & True north corr. & Platescale \\
 & & & s & & $^\circ$ & $"$ & $H$ band & & $^\circ$ & mas pixel$^{-1}$ \\
\hline        
2018-04-24 & IFS & YJH & 16 & 174 & 53.0 & 0.35 & 0.91 & 1.03 & $-1.761 \pm 0.06$ & $7.46 \pm 0.02$ \\
2018-04-24 & IRDIS & K12 & 16 & 104 & 43.8 & 0.35 & 0.91 & 1.03 & $-1.761 \pm 0.06$ & $12.256 \pm 0.016$ \\
2021-06-04 & IFS & YJH & 16 & 186 & 53.2 & 0.33 & 0.90 & 1.03 & $-1.780 \pm 0.07$ & $7.46 \pm 0.02$ \\
2021-06-04 & IRDIS & K12 & 16 & 161 & 51.3 & 0.33 & 0.90 & 1.03 & $-1.780 \pm 0.07$ & $12.261 \pm 0.005$ \\
\hline          
\end{tabular}
\end{table*}

\section{Data reduction}
\label{sec:data_reduction}

The raw data were reduced at the SPHERE Data Center by means of the SpeCal pipeline \citep{2017sf2a.conf..347D,2018A&A...615A..92G}. For IRDIS data, the standard Template Locally Optimized Combination of Images (TLOCI) reduction technique \citep{2007ApJ...660..770L} was employed, as in previous BEAST publications \citep{2021A&A...646A.164J,2021Natur.600..231J} and an additional PACO simultaneous Angular and Spectral Differential Imaging (ASDI) analysis \citep{2020A&A...637A...9F} was carried on at a later stage only with regard to one specific companion candidate (Section~\ref{sec:cc0}). For IFS we used three data analysis methods: ASDI and monochromatic Principal Component Analysis (PCA) \citep{2015A&A...576A.121M}, PACO ASDI \citep{2020A&A...637A...9F} and TRAP \citep{2021A&A...646A..24S}. The calibration of true north and pixel scale employs observations of far compact stellar cluster as in \citep{2016SPIE.9908E..34M}, while the recovery of astrometry and photometry of the companion candidates relied on the injection of negative planets on their position and minimization of residuals.

\begin{figure*}
\centering
\includegraphics[width=\hsize]{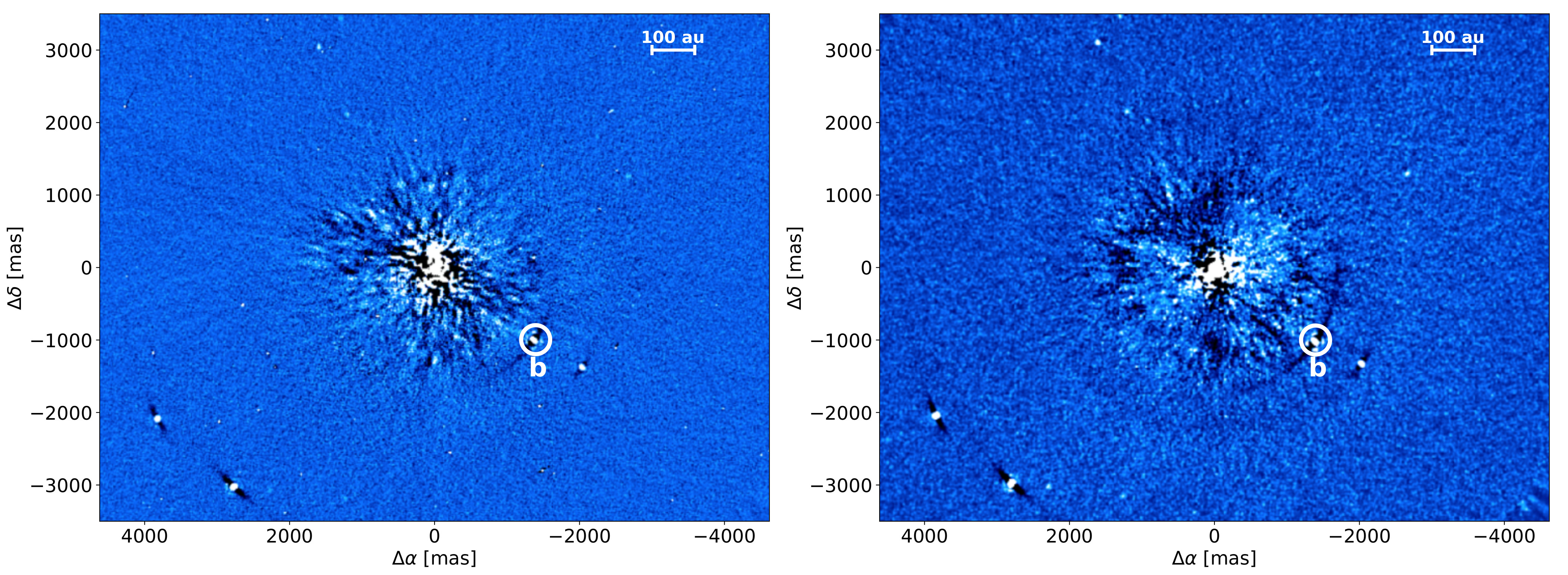}
   \caption{IRDIS images of $\mu^2$ Sco obtained with TLOCI ({\it left panel}: first epoch; {\it right panel}: second epoch). The star, artificially obscured by the coronagraphic mask, is at the center of the image. Several background sources can be easily seen as bright point sources. $\mu^2$ Sco b is the source inside the white circle.}
\label{fig:IRDIS_FoV}
\end{figure*}

\begin{figure*}
\centering
\includegraphics[width=\hsize]{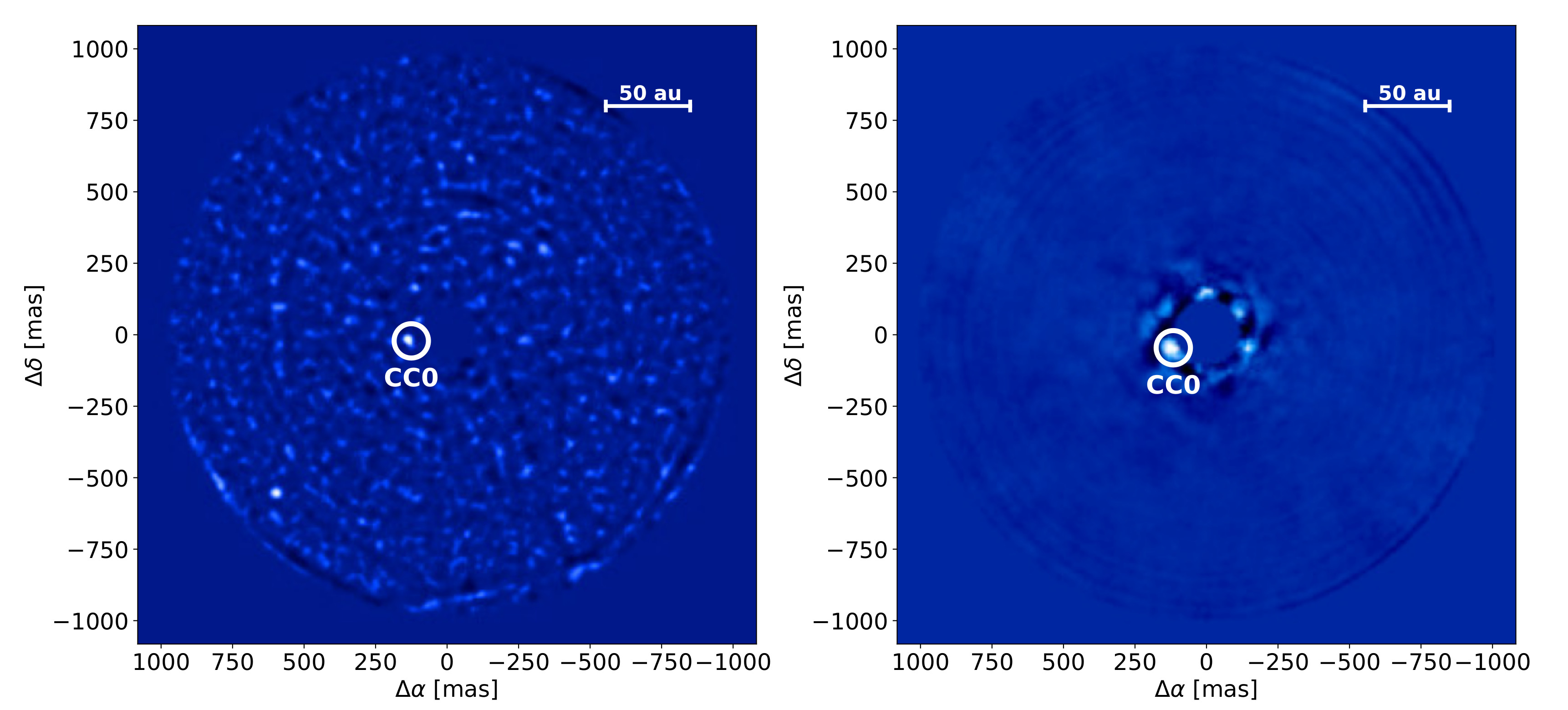}
\caption{IFS images of $\mu^2$ Sco obtained with ASDI-PCA with 25 modes ({\it left panel}: first epoch; {\it right panel}: second epoch). The star is at the center of the image but not visible due to the presence of the coronagraphic mask --which artificially masks the innermost $\sim 90$ mas of the image-- and the aggressive data analysis method used. The probable companion CC0 is the source inside the white circle. A background source is visible on the lower left of the first-epoch image.}
\label{fig:IFS_FoV}
\end{figure*}

The final IRDIS and IFS images are shown in Figure~\ref{fig:IRDIS_FoV} and Figure~\ref{fig:IFS_FoV}, respectively.
    
\section{Companion candidates}
\label{sec:cc_candidates}

\subsection{Identification of companion candidates}
\label{sec:all_ccs}

Two companion candidates (CCs) were identified inside the FoV of SPHERE’s Integral Field Spectrograph (IFS), while 34 sources were identified inside the wider field of the Infra-Red Dual-beam Imaging and Spectroscopy (IRDIS). The quality of the final IRDIS image is somewhat better for the second epoch, which is on average $\sim 0.25$ mag deeper than the first one: this explains why some of the faintest CCs in the IRDIS FoV are only detected in the second epoch. The derived properties of all the CCs within the IFS and IRDIS FoV, labeled through a unique numeric ID, can be found in Table~\ref{table:cc_info}.

\subsection{The closest companion candidate: CC0}
\label{sec:cc0}

Unlike the other CCs, the proximity of CC0 to the edge of the coronagraphic mask and its faintness invokes special caution to rule out the possibility of a systematic artifact. In order to assess the reliability of the detection of CC0, the analysis of IFS data was performed -- as already mentioned above -- by means of three different methods:
\begin{itemize}
    \item ASDI-PCA consists in the subtraction of the first $n$ principal components from the 4D data cube, obtained by combining the 39 individual monochromatic images -- corresponding to the different DITs on the detector -- after radially scaling them with wavelength; to avoid the data cube to be too large, we averaged each consecutive DIT; after dropping a few poor images, the final data cube includes $90\times39=3510$ images. We considered $n=25$ modes, a value suited for small separations, and averaged the resulting data cube according to the expected contrast spectrum for a late M star. We then constructed a signal-to-noise ratio (S/N) map where the noise is evaluated along rings at constant separation and corrected it for the low number statistics penalty factor given by \citet{2014ApJ...792...97M}. In this way, we detect CC0 in the first epoch as a source with peak S/N $\sim$ 8.0, $\text{separation} = 128 \pm 2$ mas and $\text{PA}=99.5 \pm 1.1^\circ$. Although not significant (S/N=2.5), a source at a similar location ($\text{separation}=123 \pm 7$ mas, $\text{PA} = 114 \pm 4^\circ$) is tentatively spotted in the second epoch too.
    \item PACO-ASDI is based on local learning of patch covariances, in order to capture the spectral and temporal fluctuations of background structures; the statistical modeling is exploited to provide a detection algorithm and a spectrum estimation method. The modeling of spectral correlations is useful in reducing detection artifacts. Using this algorithm, we detected the same source found with ASDI-PCA in both epochs but with higher S/N: S/N=10.6, $\text{separation} \sim 126 \pm 2$, $\text{PA}=99.9 \pm 0.8^\circ$ for the first epoch, S/N=5.0, $\text{separation}=125 \pm 3$, $\text{PA}=111.0 \pm 1.6^\circ$ for the second epoch. These S/N estimation include the contribution of spectral correlations \citep{2020A&A...637A...9F} and are performed locally (at a scale of a few pixels). This method tends to be more efficient in detecting very faint signals but it also tends to produce detection maps with slightly higher residual correlations. We also checked the detection maps obtained with the less local (at a scale of a few dozen pixels) algorithm, given S/N slightly above 5 for 2018 data with and without priors. The distribution of both detection criterion is (approximately) Gaussian in the absence of sources. While the first epoch detection employed a flat spectrum (i.e., no prior) and a late-M spectral template, the second epoch detection was obtained only when using the latter prior;
    \item TRAP is a data-driven approach to modeling the temporal behavior of stellar light contamination, rather than the spatial distribution of the speckle halo. Due to the field-of-view rotation of the image sequence, each pixel affected by planet signal sees a distinctive light curve which can be modeled and thereby distinguished from other systematic temporal trends of the data. This is achieved by a causal regression model, trained on other pixel light curves at a similar separation from the host star and fit simultaneously with the planet model for each pixel. This approach works well at small separations: no training data is lost due to insufficient field rotation, as the model is not trained locally. Default parameters, as described in the original paper, were used. The resulting contrast maps for each channel were then fitted with a L-type, T-type and flat contrast spectral template, assuming a BT-NextGen stellar model \citep{2011ASPC..448...91A} of the companion. In addition we used the extracted PACO-ASDI spectrum of CC0 as a template to derive contrast limits for the object. Given the measured contrast by PACO-ASDI, a S/N between 2.5 and 3.5 would be expected at the position, but no signal above S/N=0.5 was detected.
\end{itemize}
    
Assuming Gaussian noise distributions and considering the number of independent points sampled in an IFS image ($\sim 2\times 10^4$), the S/N obtained with PCA-ASDI and PACO-ASDI corresponds to values of the False Alarm Probability (FAP) equal to $\sim 10^{-11}$, that are extremely low even accounting for the fact that we have observed at least once 75 stars so far in the BEAST survey. However, these values underestimate the real FAP because the distributions are actually not exactly Gaussian\footnote{We ran the Shapiro-Wilk test for Gaussianity (using the on-line calculator available at \url{https://www.statskingdom.com/shapiro-wilk-test-calculator.html}) for the inner 50$\times$50 pixels (about 0.37$\times$0.37 arcsec) region of all these S/N map, eliminating the area within 0.5~$\lambda$/D of the coronagraphic edge and that around CC0 and using pixels having even indices in both coordinates, to avoid the concern related to covariance of adjacent pixels. Depending on the case, this test found or not some small deviations from Gaussianity. However, the kurtosis is almost always higher than expected for a Gaussian distribution, suggesting that outliers are more common than expected with a Gaussian distribution.} (likely because of the edge effects introduced by the coronagraphic mask) and the FAP is therefore higher and not well determined.

The physical nature of CC0 is reinforced by the observation that, at both epochs, CC0 is the peak of the IFS S/N map, and that the second-epoch PA is found within $\sim 11^\circ$ from the first-epoch PA, the separation being the same ($d_{CC0}\approx 127$ mas) within errors. The probability $p_{RPA}$ of this fact happening by chance was estimated in an empirical way. We expect the astrometric shift between the two epochs $\Delta r$ not to be larger than the fraction $\Delta r_{max}$ of the face-on circular orbit with $r=d_{CC0}$ covered over the timespan $\Delta t \approx 3.1$ yr separating the two epochs:
\begin{equation}
\Delta r \lesssim \Delta r_{max} = \frac{2 \pi d_{CC0}}{T_{orb}} \cdot \Delta t \approx 78 \text{ mas},    
\end{equation}
where $T_{orb}=1 \text{ yr} \cdot (s$ [au]$)^{3/2} \cdot (M_* [M_\odot])^{-1/2}) = 1 \text{ yr} \cdot \sqrt{21^3/9.1} \approx 32$ yr is the period of the circular orbit with $s=1 \text{ au} \cdot d_{CC0}/\text{parallax} \approx 21$ au around $\mu^2$ Sco ($M_*=9.1 M_\odot$). As a comparison, the observed shift of CC0 is $\Delta r_{obs}=26 \pm 4$ mas.

Due to the presence of the coronagraph at $\sim 100$ mas, we conveniently define as "interesting area" the annular sector, centered on CC0's first-epoch PA, with separation $100 \text{ mas} < d < 200 \text{ mas}$ and semiangular width $\Delta \text{PA}= \frac{\Delta r_{max}}{2\pi d_{CC0}} \cdot 360^\circ \approx 35^\circ$.

We then examined all BEAST images and verified that --after deleting those with clear companion candidates-- in 19 out of 99 IFS images the S/N peak lies within 200 mas from the central star. The probability $p_{RPA}$ of a nonphysical peak of the S/N map (e.g., a speckle) to be found within the interesting area defined above is given by $p_{RPA} \sim 19/99 \cdot 35^\circ/360^\circ \sim 4\%$.

\begin{figure*}
\centering
\includegraphics[width=\hsize]{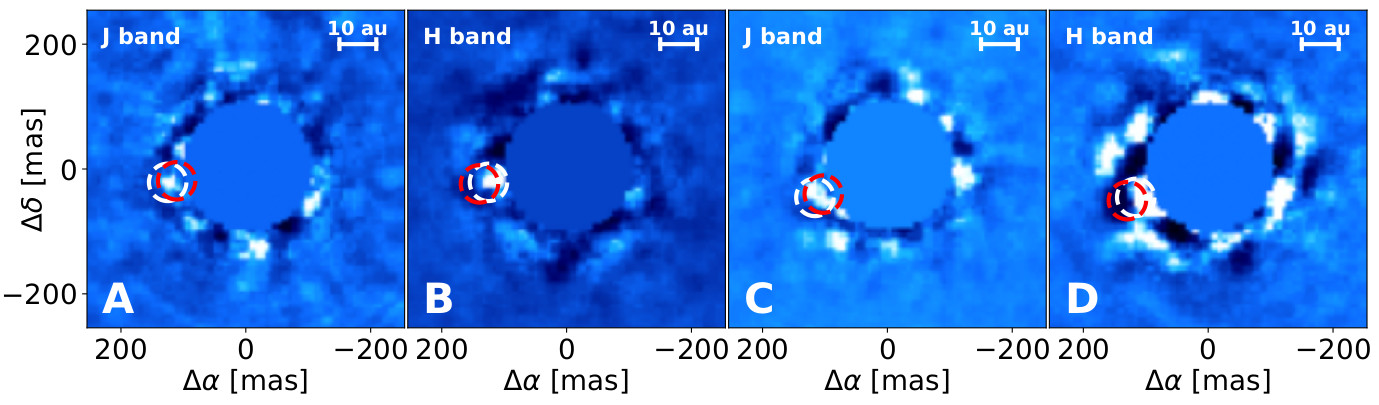}
   \caption{Position of CC0 at different wavelengths. (A) J-band image, first epoch; (B) H-band image, first epoch; (C) J-band image, second epoch; (B) H-band image, second epoch. The Y-band image is not shown due to the extreme faintness of the source. White circles indicate the best-fit positions from Table~\ref{table:cc_info}; circle radii have been enlarged to $\sim 10$ times the uncertainties on best-fit positions to enhance visibility. Unlike a speckle (red circle), the separation of CC0 does not increase with wavelength.}
\label{fig:l_shift}
\end{figure*}

Finally, the separation of CC0 is constant with wavelength at both epochs (Figure~\ref{fig:l_shift}), further supporting the detection, since separation of speckles from the center of the image is expected to be proportional to wavelength. Intriguingly, a bright spot exactly at the location of CC0 is also detected with TLOCI in IRDIS $K_1$\ and $K_2$\ data at first epoch -- although at a very low S/N$\sim$2-- However, given the very low S/N of these IRDIS detections, we consider them as upper limits in the following discussion.

On the other hand, CC0 was not detected using TRAP. The limiting magnitude obtained with this reduction is brighter than the PCA-ASDI and PACO-ASDI detections (because the wavelength dependence of speckle position is not exploited by TRAP), but should result in a marginal detection; it is at present unclear if the nondetection is due to lacking sensitivity as no Spectral Differential Imaging is used, or because the ASDI techniques underestimate errors and false alarm probabilities. For the time being we suggest that caution must be taken about the actual detection of CC0, that we consider as a probable detection but not a definite one.

\section{Analysis of companion candidates}
\label{sec:cc_analysis}

\subsection{Companion confirmation}
\label{sec:cc_confirmation}

The confirmation of a directly-imaged companion candidate is usually performed by checking if the proper motion is similar to that of the target star and significantly different from that of the background sources, that is stars that are located far behind the target but appear projected close to it. Taking the first epoch as a reference, sources with a null proper motion are shifted in the second epoch as a reflex motion due to the target’s proper motion over the elapsed time.
    
As expected from the low galactic latitude ($b=3.86^\circ$) of the target, the final images are abundant in detected sources: two companion candidates are seen inside the IFS FoV Figure~\ref{fig:IFS_FoV}, while in the wider IRDIS FoV there are 46 objects (Figure(~\ref{fig:IRDIS_FoV}) detected either in the first or in the second epoch, with the majority -34- being observed in both epochs. Excluding CC1 -seen in both epochs by IFS- we are left with 11 dim CCs only seen by IRDIS in the second epoch.
    
\subsubsection{Photometric analysis}
\label{sec:cc_phot_analysis}
    
\begin{figure}
\centering
\includegraphics[width=\hsize]{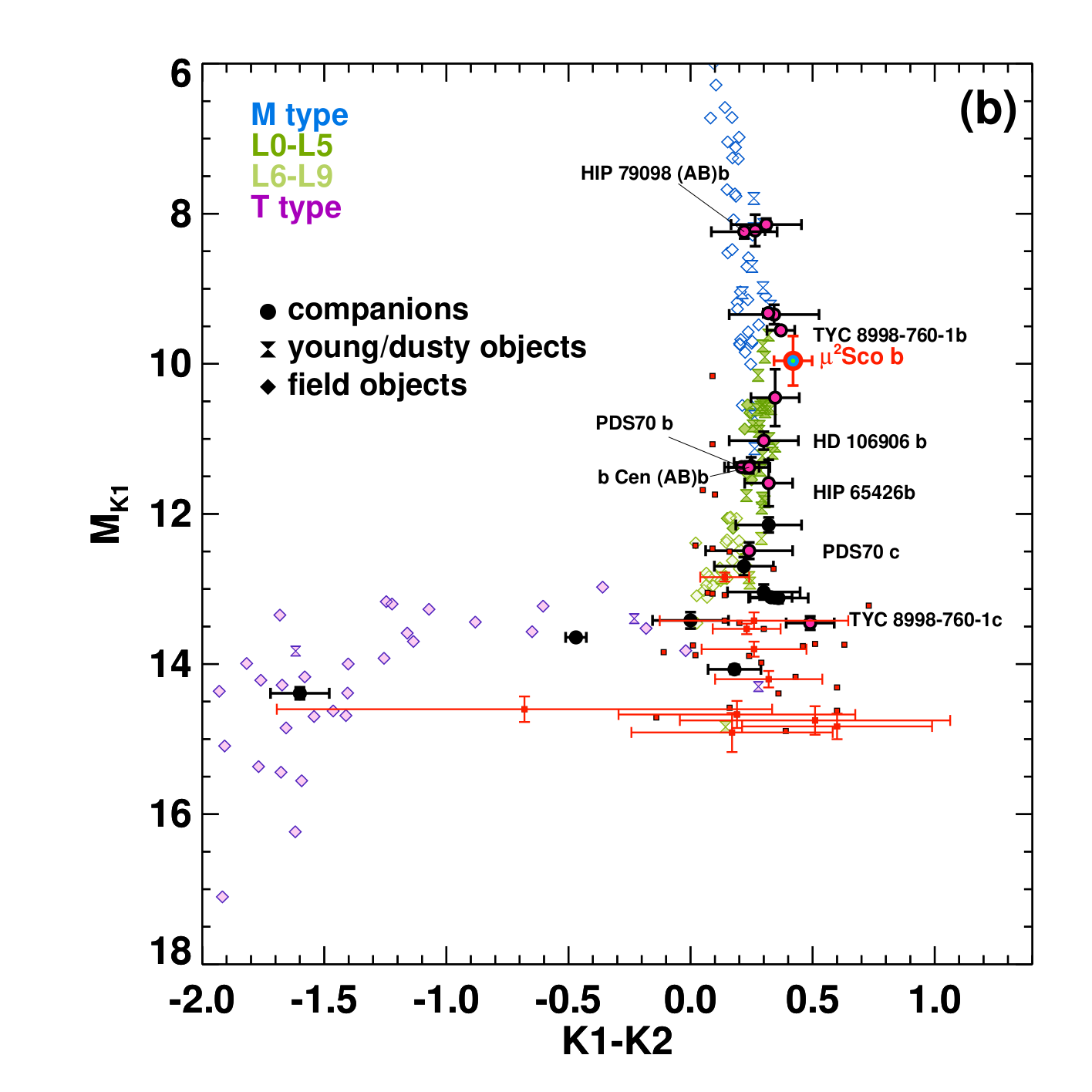}
\caption{($K_1-K_2$, $K$) color magnitude diagram of known substellar companions and field objects. IRDIS CCs only seen in one epoch are shown as red dots. The position of CC0 (=$\mu^2$ Sco b) is consistent with that expected for a young object.}
\label{fig:CMD}
\end{figure}

Before analyzing the astrometric shifts of the CCs seen at both epochs, we tried to assess whether 9 dim CCs that are only seen in the second epoch could be confidently labeled as background contaminants based on their colors (Figure~\ref{fig:CMD}). However, owing to the large color uncertainty, their position in the ($K_1-K_2$, $K_1$) color-magnitude diagram (CMD) does not provide us with a definitive answer. The same applies for CC29 and CC30, only seen in $K_1$\ in one epoch. Given that the geometric probability of a random alignment between a background source and a target star with separation $d$ is $\propto d^2$, we are only able to argue that the presence of several secure background sources that have a smaller separation to $\mu^2$ Sco than each of these 11 sources\footnote{5 sources for CC7 and CC8, 6 for CC10 and CC11, 14 for CC20, 18 for CC25, 21 for CC29 and CC30, 24 for CC34 and CC35, 30 for CC42.}, is strongly suggestive of a background nature for the 11 sources as well. We excluded all these CCs from the following analysis, in absence of any evidence for a physical association to our target.

\subsubsection{Astrometric analysis}
\label{sec:cc_astrom_analysis}

\begin{figure}
\centering
\includegraphics[width=\hsize]{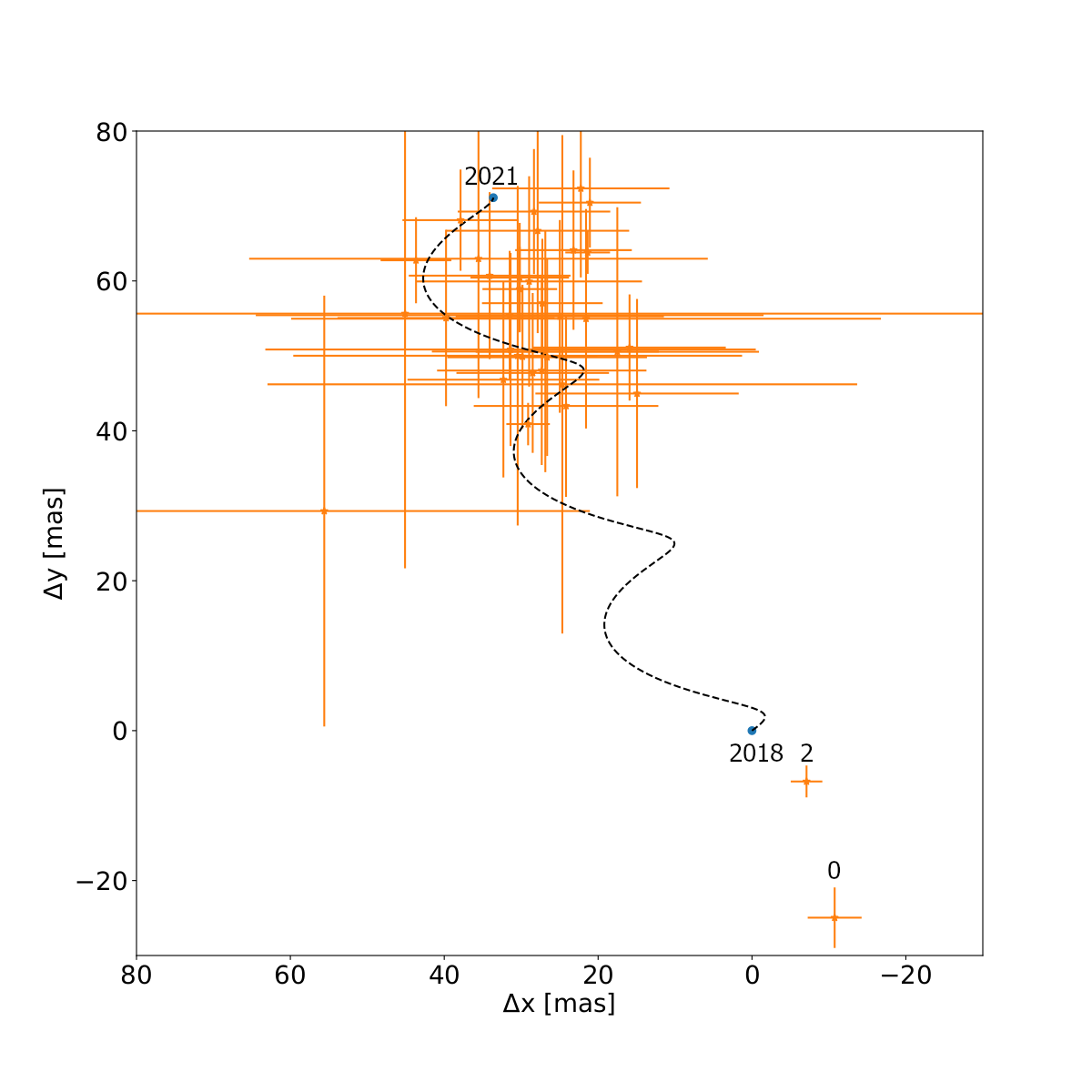}
\caption{Astrometric motion of CCs present in both epochs. As the plot shows the difference in separation between the two epochs, a comoving source should be close to the origin (labeled by "2018"). A background source with null proper motion will move according to the dashed curve as a reflection of the target’s motion, ending in the blue circle labeled as ‘2021’ at the second epoch. The motion of CC0 (labeled by '0') and CC2 (labeled by '2') is distinct from the cloud of background contaminants.}
\label{fig:astro_shifts}
\end{figure}

With regard to the IRDIS CCs seen in both epochs, we are able to confidently label 33 out of 34 objects as background interlopers through astrometric analysis (see Figure~\ref{fig:astro_shifts}).

On the other hand, CC0 and CC2 show a completely different motion compared to the remaining sources. The mean astrometric shifts of this cloud of sources with respect to the position expected for a static source are $-5.5\pm 1.2$~mas (rms: 6.7 mas) along right ascension, $-15.2\pm 1.8$~mas (rms: 10.2 mas) along declination. The astrometric shift of CC0 is $-44.4\pm 3.5$~mas along RA, $-96.1\pm 4.0$~mas along declination; that of CC2 is $-40.7\pm 2.1$~mas along RA, $-77.9\pm 2.1$~mas along declination. Hence, the observed shifts of CC0 and CC2 are $\sim 8 \sigma$ and $\sim 9 \sigma$ away from the cloud of background sources (we call it "CC cloud"). The conclusion is robust to systematic astrometric offsets between the two epochs, equally affecting all the CCs. CC0 and CC2 are then plausible candidates as physical companion. We examine their cases in more detail in Section~\ref{sec:fap}.

\subsubsection{Motion of background sources}
\label{sec:cc_motion}

As shown by Figure~\ref{fig:astro_shifts}, the CC cloud is shifted on average by $\sim -16$ mas with respect to the position expected on the basis of the motion of $\mu^2$ Sco. After carefully checking the centering of our images to exclude a large systematic effect, we are left with three possible causes (or a combination of them):
\begin{enumerate}
    \item the adopted proper motion is not correct, the error being approximately 3 mas/yr in both right ascension and declination;
    \item the photocenter of $\mu^2$ Sco is offset with respect to the barycenter of the system by a different quantity at the two epochs. This might or might not be related to point 1;
    \item the field stars have on average a proper motion that is not null. This might be a reflection of the galactic rotation curve (it is important to note that the line of sight toward $\mu^2$~Sco passes at about 15 degrees from the galactic center and the background interlopers are expected to be stars of the inner part of the galactic disk at a few kpc from the Sun).
\end{enumerate}

Scenario 1 seems unlikely: the proper motion considered for $\mu^2$ Sco is very similar to that of its neighbor $\mu^1$ Sco ($\mu_\alpha=-11.867 \pm 0.043$ mas yr$^{-1}$ in RA and $\mu_\delta=-22.511 \pm 0.035$ mas yr$^{-1}$ in Dec) and to the average value for ELS ($\mu_\alpha=-11.57 \pm 0.48$ mas yr$^{-1}$ and $\mu_\delta =-22.12 \pm 0.70$ mas yr$^{-1}$); the differences are too small to explain the observed residuals.

If the correct scenario were 2, it would apply to both field stars and physical companions; indeed, the motion of CC0 and CC2 appears similar to the offset of the field stars with respect to the prediction for null motion. This would require a substantial motion of $\mu^2$ Sco caused by an (unseen) companion with an orbital period less than 24 yr; a massive short-period companion would likely cause large variations in the radial velocities, which instead are rather constant within a few hundreds of m/s over $\sim$ 10 years (see Appendix~\ref{sec:binarity}).

To investigate scenario 3, we looked for Gaia EDR3 data of field stars near $\mu^2$ Sco. Given that none of the IRDIS background stars is bright enough to be present in Gaia, we searched for field stars within 5 arcmin from our target (we label it "Gaia bg sample"). The main properties of the Gaia bg sample, composed of 6286 stars, are:
\begin{eqnarray*}
        <G>&=&20.09~\text{mag,} \\
        <\pi>&=&0.30~\text{mas,} \\
        <\mu_\alpha^*>&=&-2.681 \pm 0.042~\text{ mas yr}^{-1} ~(rms: 3.367~\text{ mas yr}^{-1}), \\
        <\mu_\delta>&=&-3.848 \pm 0.043~\text{ mas yr}^{-1} ~(rms: 3.431~\text{ mas yr}^{-1}).
\end{eqnarray*}
    
The mean proper motion of the CC cloud, that is simply the ratio between the mean astrometric shift and the time baseline, is instead $-1.77 \pm 0.37$ mas yr$^{-1}$ (rms: 2.17 mas yr$^{-1}$) along RA and  $-4.87 \pm 0.56$ mas yr$^{-1}$ (rms: 3.29 mas yr$^{-1}$) along dec. We anticipate that a small systematic offset, of about $-0.6 \pm 0.4$ mas along RA and $+1.7 \pm 0.4$ mas along dec, is likely to affect the proper motions of our CCs (see Section~\ref{sec:fap_bg}). With this caveat in mind, we see that the proper motion of the Gaia bg sample is indeed similar to that of the CC cloud; the latter is likely the M-type part of the same population probed, at brighter magnitudes, by the former, as indicated by the following qualitative argument. From the median parallax of the Gaia bg sample we infer a distance of $\sim$3 kpc, and absolute magnitudes $M_G \sim 7.5$, that correspond to late K stars; since we are neglecting reddening, these objects may be even intrinsically brighter. This G magnitude translates to $M_K \sim 4.8$ according to the tables by \citet{2013ApJS..208....9P}. The corresponding apparent magnitude $K \sim 17.4$ gives a contrast with respect to $\mu^2$ Sco of d$K=13.1$ mag; these stars are roughly two magnitudes brighter than the CC cloud (median contrast: d$K \sim 15$ mag). If we assume that IRDIS CCs are on average at the same distance of the Gaia bg sample, their absolute $K$\ magnitudes are $M_K \sim 6.7$; according to the tables by \citet{2013ApJS..208....9P} this corresponds to M3V stars that have an absolute $M_G \sim 10.0$, and to an apparent magnitude $G \sim 22.6$. IRDIS CCs are too faint to be detected by Gaia (that has a limiting magnitude of $G \sim 21$), and conversely we expect about a couple of Gaia stars (with a density of 1 star every 45 square arcsec) within the IRDIS FoV; the relative frequency agrees with expectation for a reasonable mass distribution (e.g., Salpeter-like) if they belong to the same parent population.

\subsection{Confirmation of physical companion(s) to \texorpdfstring{$\mu^2$}{mu2} Sco}
\label{sec:fap}

In this subsection, we examine the possibility that CC2 is not a bound companion of $\mu^2$ Sco considering two possible alternatives:
\begin{itemize}
    \item that CC2 is a high proper motion background star;
    \item that CC2 is a brown dwarf (BD) member of Scorpius-Centaurus that appears projected close to $\mu^2$ Sco.
\end{itemize}

We note that while the following considerations refer to CC2, they equally apply to CC0 which has a higher proper motion and a smaller separation than CC2, hence intrinsically lower false alarm probabilities (FAP).

\subsubsection{CC2 is not a background star}
\label{sec:fap_bg}

Having shown that the cloud of sources seen in the IRDIS images is made of background interlopers, it is necessary to estimate the probability that an object drawn from the same population could have been considered an "interesting" companion candidate to $\mu^2$ Sco because of a "remarkable" astrometric shift relative to $\mu^2$ Sco. Two factors must be taken into account: the probability of finding an object within certain boundaries of $\mu_\alpha^*$ and $\mu_\delta$; the fact that we have fully\footnote{That is, at least twice. As already mentioned, two observations are needed to compute relative proper motions.} observed 25 stars in the survey.

Starting from the Gaia bg sample, we identify as interesting the stars with $\mu_\alpha^*<-9.8$ mas yr$^{-1}$ and $\mu_\delta <-21.1$ mas yr$^{-1}$ \footnote{This is a conservative approach that considers interesting all the objects having a larger relative proper motion -- with respect to the CC cloud -- than $\mu^2$ Sco itself; see below.}. After excluding a few objects that, despite passing the test, have parallax $>5$~mas (which would correspond to $4 \lesssim dK \lesssim 10$ mag) and are thus so bright that they would have been disguised as stellar -- and not substellar -- companions, we obtain that 5 out of 6286 stars satisfy this criterion. This implies a fraction of interesting background objects of $f_G=1.67 \cdot 10^{-5}$. The FAP of finding one star with these features in our observations of $\mu^2$ Sco (that is, within the IRDIS field of view), given that we observe 36 CCs in both epochs, is then $6 \cdot 10^{-4}$. 
    
A similar argument can be made by creating a synthetic sample of background stars by means of the Besan\c{c}on Galaxy Model interface\footnote{Available at \href{https://model.obs-besancon.fr/modele\_home.php}{https://model.obs-besancon.fr/modele\_home.php}.} \citep{2014A&A...564A.102C}. We selected a sample of stars with distance $d \in [0, 50]$ kpc, apparent $K$ magnitude similar to that observed for CC0 ($K \in [15, 17]$ mag), radius $\rho=5$~arcmin around the position of $\mu^2$ Sco. Out of 6595 stars, just 4 (corresponding to $f_B=1.34 \cdot 10^{-5}$) pass the proper motion test described above; this yields a FAP -given 36 CCs- of $4.8 \cdot 10^{-4}$.

Thus, the Besan\c{c}on-based test and the Gaia-based test give very similar results. We point out that, by defining an unbound region of the proper motion space of ‘fast-moving stars’, we are actually overestimating the FAP: a source with an unusually large absolute value of the proper motion would not have been consistent with being bound to the star, and would have been discarded.
    
The probability of having seen at least one background source with these features throughout the entire survey can be estimated in this way. The median proper motion of BEAST targets is -in absolute value- larger than that of $\mu^2$ Sco, but we assume for simplicity a strict equality; if we further assume that the fractions $f_B$ and $f_G$ computed above do not depend on the sky coordinates, we can extend the above reasoning to the whole survey. Given that we have fully observed 25 targets, and that on average we see $\sim 11$ sources per observation \citep{2021A&A...646A.164J}, we obtain a FAP of having seen at least one background object disguised as a possible companion within the entire survey of $4.6 \cdot 10^{-3}$ (Gaia-based test) and $3.7 \cdot 10^{-3}$ (Besan\c{c}on-based test). We conclude that CC2 (and even more likely CC0) is not a background object at a high level of confidence.

\subsubsection{CC2 is not a free-floating brown dwarf}
\label{sec:fap_bd}

Whilst confirming that the sources are comoving, the astrometric analysis does not strictly allow exclusion of an alternative scenario: namely, that the sources are free-floating UCL substellar objects that happen by chance to be close to the line of sight of $\mu^2$ Sco. In order to quantify this false alarm probability, three points need to be considered: the probability of finding an "interesting" ($M=1-75~\mj$) object; the probability of finding it within the IRDIS field of view; the fact that we have fully observed 25 stars in the survey.

Assessing the number of free-floating UCL members necessarily requires some assumptions on the initial mass function (IMF) of the association\footnote{Given that any Gaia-based census of association members is not complete at such low masses.}. 

Recently, \citet{2022NatAs...6...89M} have uncovered a rich population of free-floating planets and brown dwarfs in Upper Scorpius, extending the IMF of the association down to $0.005~M_\odot$. As US is a subregion of Sco-Cen, we expect that this IMF can be safely adopted for UCL too. By normalizing their IMF we obtain a probability density function (PDF); we are thus able to compute the fraction of objects in the mass range $[5 \mj, 75 \mj]$. We do not consider objects below $5 \mj$ not only because they are not covered by their data, but also, more importantly, because this is approximately the lower mass to which we are sensitive in BEAST. Integration of the PDF yields 0.205, meaning that one out of five objects in Sco-Cen is expected to belong to this mass range. This PDF should be multiplied by the projected density of objects in the region around $\mu^2$ Sco. To estimate this quantity, we take again the list of bona fide members compiled by \citet{2019A&A...623A.112D}. The faintest members of UCL have apparent $G=$18.5-19.5, and the corresponding masses reach down to $0.013~M_\odot$ (see “Stellar system analysis”). Since Gaia is complete within this magnitude range (its limiting magnitude is $G\sim 21$), we assume a sharp transition between 100\% completeness above 15 $\mj$ to 0\% completeness below 15 $\mj$. We rescale the number of sources we see in UCL (3842), ELS (575) and LS (73) by dividing by the integral of the PDF above 0.015 $M_\odot$, obtaining a complete census of approximately $\sim 4021$ sources for UCL, $\sim 602$ for LS, $\sim 76$ for ELS.
    
Multiplying these numbers by the integral of the PDF from $5 \mj$ to $75 \mj$, we obtain the expected number of interesting objects $\tilde{N}$:
\begin{eqnarray*}
        \tilde{N}_{LS} &\sim& 123, \\
        \tilde{N}_{ELS} &\sim& 16, \\
        \tilde{N}_{UCL} &\sim& 823.
\end{eqnarray*}
    
In order to turn the expected number of objects into a projected density (i.e. the number of expected BD interlopers per arcsec$^2$), the areas of LS, ELS and UCL must be evaluated. For UCL, coordinate boundaries $(l,b) =  [313^\circ,343^\circ] \times [2^\circ, 28^\circ]$ as in \citet{2016MNRAS.461..794P} were used. For LS and ELS we computed the mean <l>, <b> and the related standard deviations $\sigma_l$ and $\sigma_b$, and defined the areas as $[<l>-2\sigma_l, <l>+2\sigma_l] \times [<b>-2\sigma_b, <b>+2\sigma_b]$. We get:
\begin{eqnarray*}
        A_{LS} &=& 23.2~\text{deg}^2=3.0 \cdot 10^8~\text{arcsec}^2, \\
        A_{ELS} &=& 3.8~\text{deg}^2=4.9 \cdot 10^7~\text{arcsec}^2, \\
        A_{UCL} &=& 747.0~\text{deg}^2=9.7 \cdot 10^9~\text{arcsec}^2,
\end{eqnarray*}
    
\noindent so that the projected densities are:
\begin{eqnarray*}
        \Sigma_{LS} &=& \tilde{N}_{LS}/A_{LS} = 4 \cdot 10^{-7}~\text{arcsec}^2, \\
        \Sigma_{ELS} &=& \tilde{N}_{ELS}/A_{ELS} = 3.2 \cdot 10^{-7}~\text{arcsec}^2, \\
        \Sigma_{UCL} &=& \tilde{N}_{UCL}/A_{UCL} = 8.5 \cdot 10^{-8}~\text{arcsec}^2.
\end{eqnarray*}

The mean density of sources in the environment of BEAST stars is usually not as high as in LS. The value that is more representative of the median behavior of the sample is probably that of UCL; we retain the value for LS as a (very high) upper limit to this source of contamination. It is possible now to estimate the number of BD interlopers expected in the IRDIS FoV (a square of $11" \times 11"$):
\begin{eqnarray*}
        N_{BD,UCL} &=& \Sigma_{UCL} \cdot (11")^2 = 1.0 \cdot 10^{-5}, \\
        N_{BD,LS} &=& \Sigma_{LS} \cdot (11")^2 = 5.0  \cdot 10^{-5},
\end{eqnarray*}
and finally the false alarm probability of having seen at least one of these objects across the 25 targets that we have observed at least twice using a binomial distribution is:
\begin{eqnarray*}
        p_{BD,UCL} &=& 2.6 \cdot 10^{-4}, \\
        p_{BD,LS} &=& 1.0 \cdot 10^{-3}.
\end{eqnarray*}
The probability of a chance alignment by free-floating objects is negligible for CC2 (and even more for CC0).

Finally, the position of both CC0 and CC2 is constant with wavelength, as expected for physical sources. We are therefore able to confirm in a robust way the source CC2 as being a substellar companion to $\mu^2$ Sco ($\mu^2$ Sco b). As regards CC0, based on the reliability arguments presented in the Section~\ref{sec:cc0}, we consider it to be a probable detection that would make it a second physical companion to $\mu^2$ Sco.

\subsection{Companion characterization}
\label{sec:cc_characterization}

The star $\mu^2$ Sco appears to be surrounded by one, and possibly two, physical companions. We recall that, while the existence of CC2 is firmly established (S/N=80–120), the detection of CC0, owing to its extreme proximity to the target, is more sensitive to the reduction method employed, stretching from a robust detection (S/N=10.6 in the first epoch, S/N=5.7 in the second epoch) in one case to a nondetection where a marginal detection (S/N=2.5–3.5) was expected in another case. 

Given the not unambiguous outcome of the different reductions and the subtleties in the derivation of confidence levels and false alarm probabilities, we highlight the need for follow-up observations to definitely confirm or rule out its existence. With this caveat in mind, throughout the text, we continue to refer to both the robust $\mu^2$ Sco b and the probable CC0, deriving the properties of the latter from the results of our successful detections.

\subsubsection{Spectra and photometry}
\label{sec:spectra}

The PACO-ASDI algorithm \citep{2018A&A...618A.138F} provides a spectrum for the probable companion CC0 for both epochs. As expected, the spectrum relative to the second epoch is characterized by a significantly lower S/N than the first-epoch one\footnote{As a consistency check, we stress that in the H-band region, where the signal is highest, the two spectra are consistent with one another.}. From this moment on we always refer to the first-epoch spectrum, the only one which allows a tentative characterization of CC0.

\begin{figure}
\centering
\includegraphics[width=\hsize]{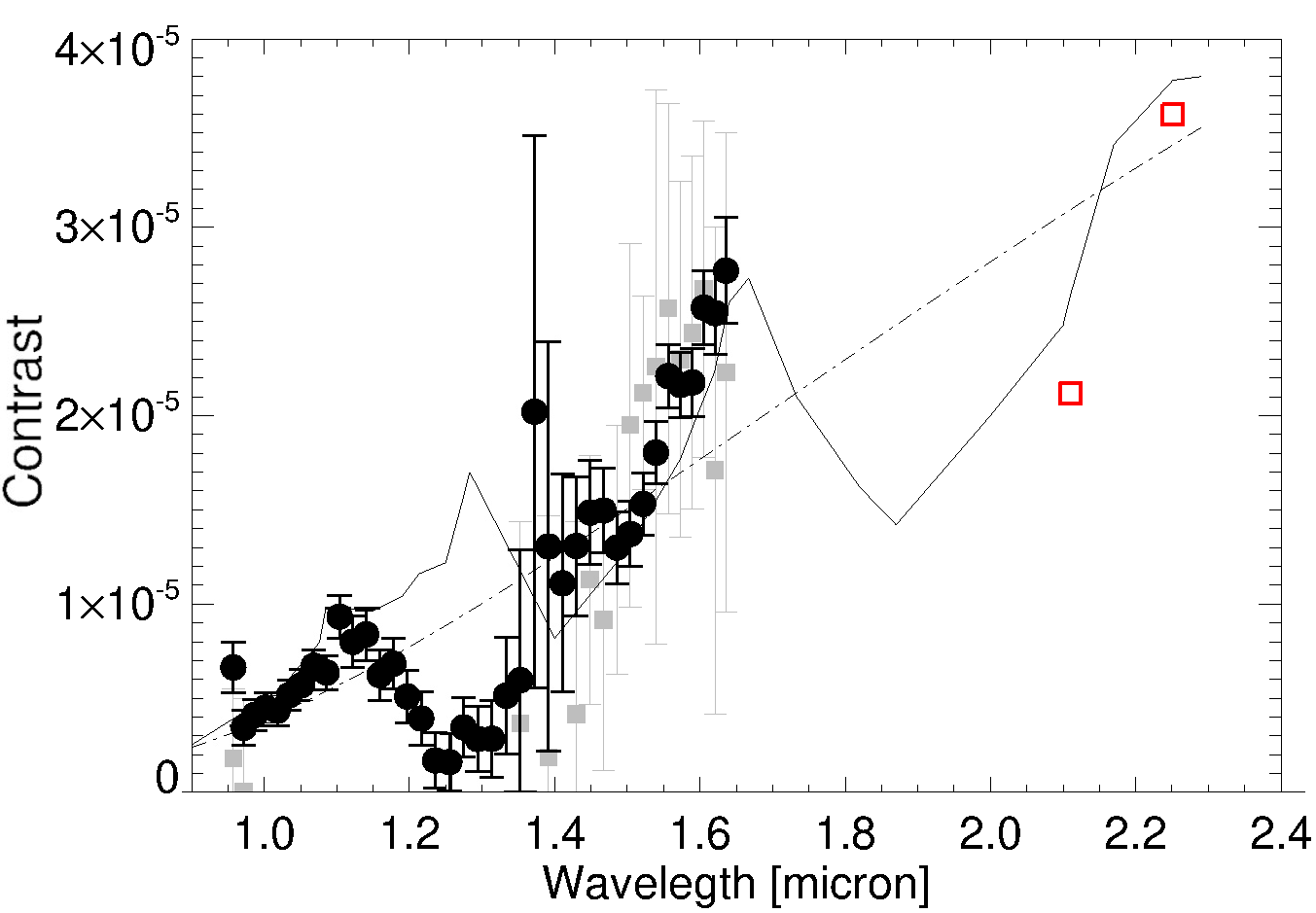}
\caption{Contrast spectrum for CC0 obtained with PACO-ASDI from 2018 IFS data (black dots) and from PACO-ASDI for IRDIS data acquired at the same epoch ($5\sigma$ upper limits; red squares). The second-epoch ASDI-PCA spectrum (gray squares) is shown for reference. The solid line is the contrast spectrum expected for a 20 Myr old, $19.5 \mj$ brown dwarf using AMES-Dusty models. The dot-dashed line is a black body curve with the same temperature and radius as this model.}
\label{fig:c_spectrum}
\end{figure}

After combining the spectrum from the first epoch with upper limits in the $K_1$ and $K_2$ band provided by the same algorithm applied to the IRDIS first-epoch dataset (see Figure~\ref{fig:c_spectrum}), we fit the spectrum with those provided by the AMES-Dusty models \citep{2001ApJ...556..357A} -- suitable for substellar objects and low-mass stars -- assuming our age and distance estimates for $\mu^2$ Sco. The best-fit model has a mass of $M_c=19.5\pm0.9$~\mj, a temperature of \teff=$2262\pm 28$~K and $\log{L/L_\odot}=-3.08\pm 0.03$. The comparison is generally fairly good (reduced $\chi^2=2.55$); most of the contribution to $\chi^2$ is due to the region around the J-band, where the observed spectrum is much lower than expected. As no strong molecular band is expected in the region around $1.25~\mu$m according to any realistic atmospheric model \citep[compare, e.g.,][]{2021ApJ...920...85M}, we attribute this discrepancy to residual speckle noise.

A similar fit with BHAC15 models yields $M_c=17.4 \pm 0.9 \mj$, $\teff=2274 \pm 28$ K and $\log{L/L_\odot}=-3.04 \pm 0.03$. We account for theoretical uncertainties by averaging the two estimates to obtain a final mass of $M_c=18.5 \pm 1.5 \mj$.

We highlight that, despite the high level of irradiation at its expected orbital distance ($21 \pm 1$ au), radiation from the probable companion would be almost completely due to its self-luminosity rather than reflected light from the star. In fact, for an albedo similar to the Earth, the equilibrium temperature of CC0 is about 900 K; comparing it with our best-fit effective temperature of about 2270 K, we estimate that stellar irradiation contributes only about 2\% to the total luminosity of the object, a value well within observational uncertainties.

As regards $\mu^2$ Sco b, for which we could not extract a spectrum as it lies outside the IFS FoV, the available photometric information is limited to the measured contrasts in the $K_1$ and $K_2$ band; combining them with the $K$ magnitude of the primary, we derive absolute ($K_1$, $K_2$) magnitudes. No significant photometric variation is observed between the two epochs. The position of the companion in the ($K_1-K_2$, $K$) color-magnitude diagram confirms its compatibility with a substellar object lying at the very beginning of the sequence of L dwarfs (L0-L2 type; see Figure~\ref{fig:CMD}).

A comparison of the photometry with theoretical magnitudes from the same two models used for CC0 provides us with two mass estimates, which again we average to get a final mass $M_b=14.4 \pm 0.8$~\mj.

As regards CC0, the reported magnitudes are computed by collapsing IFS spectral channels 12-21 (1.159-1.333 $\mu$m) and 30-38 (1.504-1.636 $\mu$m) from the first epoch to build the two bands $J_{IFS}=1.246~\mu$m (band width=0.174~$\mu$m) and $H_{IFS}=1.570~\mu$m (band width=0.132~$\mu$m); the contrasts in the two bands are translated into absolute magnitudes by adding 2MASS $J$ and $H$ magnitudes of the primary.

\subsubsection{Astrometry and orbits}
\label{sec:orbits}

The position of $\mu^2$ Sco b and CC0 relative to the star is not exactly the same in the two epochs, showing a significant displacement of $\sim 10$ mas and $\sim 20$ mas, respectively: a possible hint of orbital motion around the star. Before analyzing this aspect quantitatively, we quantified the accuracy of the relative astrometry provided by SPHERE observations exploiting the large amount of background CCs present in the images.
    
Starting from the Gaia bg sample defined in Section~\ref{sec:cc_motion} (6286 sources), we select the 2741 sources with parallax
0.1 mas$<\pi<$5 mas (i.e., $0.2~\text{kpc}<d<10~\text{kpc}$). The mean proper motion of this "restricted Gaia bg sample" is $-2.51 \pm 0.07$ mas yr$^{-1}$ (rms=3.53 mas yr$^{-1}$) along right ascension, and $-3.82 \pm 0.07$ mas yr$^{-1}$ (rms=3.76 mas yr$^{-1}$) along declination.
    
To test whether the two samples are drawn from the same parent distribution, we performed two independent Kolmogorov-Smirnov (KS) tests (one for $\mu_\alpha^*$, one for $\mu_\delta$) at level $\alpha=0.05$. While the null hypothesis could not be rejected for $\mu_\alpha^*$ ($p=0.13 > \alpha$), a significant difference exists with respect to $\mu_\delta$ ($p=0.008 < \alpha$). In particular, the CC cloud appears to have a somewhat higher $\mu_\alpha^*$ and lower $\mu_\delta$ than the restricted Gaia bg sample. To quantify this, we identified the range of $\mu_\alpha^*$ and $\mu_\delta$ shifts to be solidly applied to the whole all CC cloud so that the KS test is passed at level $\alpha=0.05$.
    
The range of shifts needed for $\mu_\alpha^*$ is $\Delta \mu_\alpha^* \in [-1.44,+0.22]$ mas yr$^{-1}$, while for $\mu_\delta$ it is $\Delta \mu_\delta \in [0.55, 2.81]$ mas yr$^{-1}$. The mean values within these ranges are -0.61 mas yr$^{-1}$ and 1.68 mas yr$^{-1}$, respectively. Multiplying by the temporal baseline, and equally splitting the correction between the two epochs, we get $\Delta \alpha =-1.34$ mas, $\Delta \delta =3.73$ mas, which can be attributed to a not perfect centering of the star in one or both epochs (note that this is much smaller than what cited at Section~\ref{sec:fap_bg} and well within typical uncertainties of star centering in SPHERE astrometry: see \citealt{2016SPIE.9908E..34M}).
    
Instead of using these values to fix the astrometry of $\mu^2$ Sco b and CC0, we conservatively opted for treating $\Delta \alpha$ and $\Delta \delta$ as an additional source of random uncertainty on their relative astrometry, and propagate it to derive final uncertainties $\sigma_d$ and $\sigma_{\text{PA}}$ on separation and PA, respectively, that are somewhat broadened with respect to those in Table~\ref{table:cc_info}: for CC0, $(\sigma_d, \sigma_{\text{PA}})=(2.5 \text{ mas}, 2.0^\circ)$ and $(3.3 \text{ mas}, 2.2^\circ)$ for the first and second epoch, respectively; for CC2 = $\mu^2$ Sco b, $(\sigma_d, \sigma_{\text{PA}})=(3.1 \text{ mas}, 0.12^\circ)$ and $(2.8 \text{ mas}, 0.11^\circ)$ for the first and second epoch, respectively.

\begin{figure*}
\centering
\includegraphics[width=\hsize]{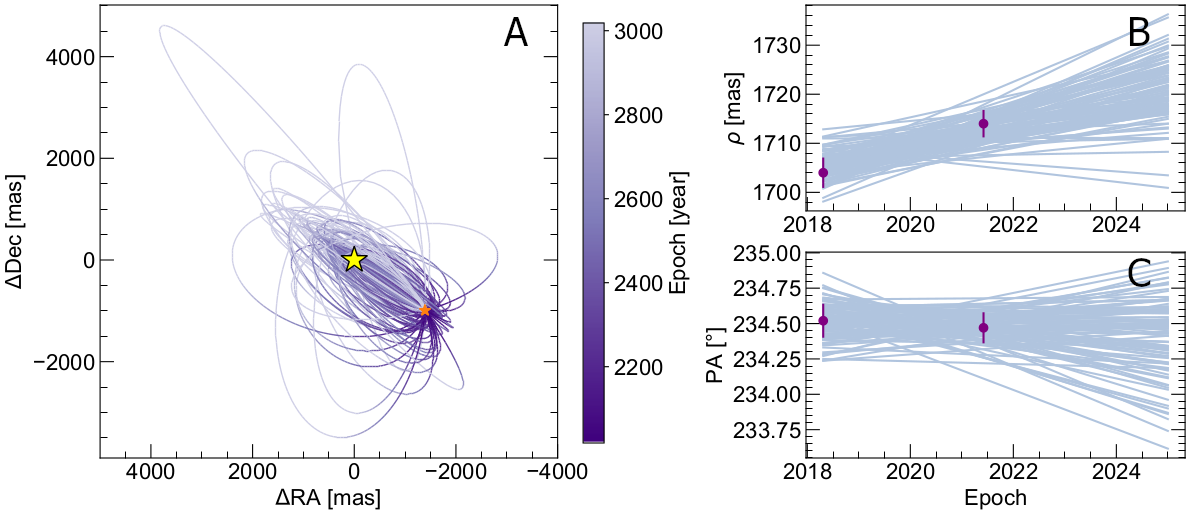}
\caption{Subset of possible orbits for the robust companion candidate $\mu^2$ Sco b. (A) Sky-projected orbital fit (orange star icon) based on the 2018 and 2021 epochs which overlap at this scale, showing 100 randomly drawn orbits from the orbitize! MCMC chains. The position of $\mu^2$ Sco is indicated by the yellow star. (B) Evolution over time of the planet-star separation ($\rho$); measured points and the corresponding error bars are shown in purple. (C) Evolution over time of the position angle (PA); again, measured points and the corresponding error bars are shown in purple.}
\label{fig:b_orbits}
\end{figure*}

\begin{figure}
\centering
\includegraphics[width=\hsize]{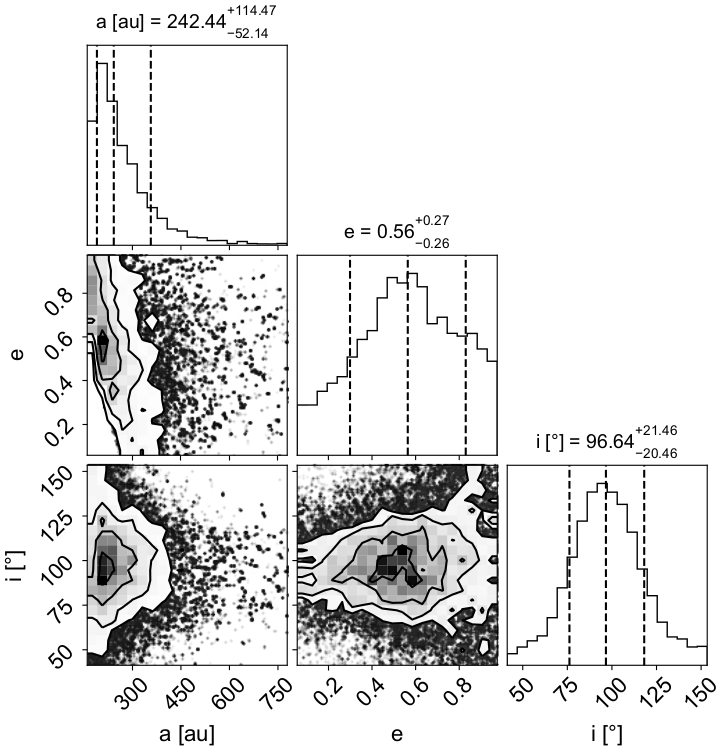}
\caption{Posterior distributions of orbital parameters for the robust companion $\mu^2$ Sco b. Corner plot showing the posterior distributions from orbitize! for semi-major axis (a), eccentricity (e) and inclination (i). The 16th (left dashed line), 50th (middle dashed line) and 84th (right dashed line) percentiles of each parameter distribution are indicated, and the 50th percentile (median) value is listed with $1 \sigma$ uncertainties derived from the lower and upper percentiles. One percent of all chains, representing long-tail values, have been excluded from the corner plots for clarity, but are still considered for the percentile calculations. No priors or constraints have been given for any of the parameters.}
\label{fig:b_param}
\end{figure}

\begin{figure*}
\centering
\includegraphics[width=\hsize]{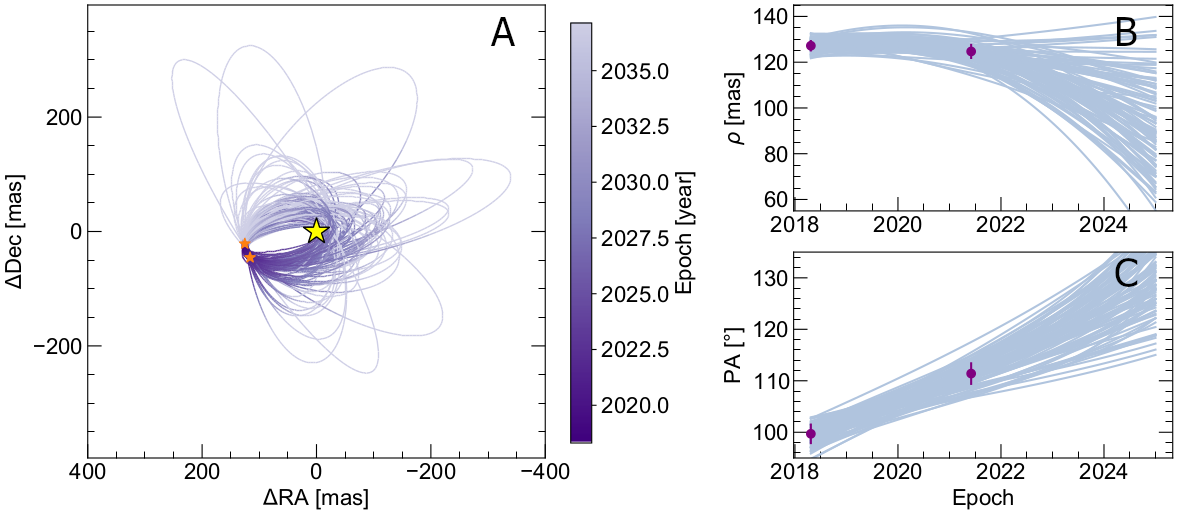}
\caption{Subset of possible orbits for the probable CC0. (A) Sky-projected orbital fit (orange star icons) based on the 2018 and 2021 epochs, showing 100 randomly drawn orbits from the orbitize! MCMC chains. The position of $\mu^2$ Sco is indicated by the yellow star. (B) Evolution over time of the planet-star separation ($\rho$); measured points and the corresponding error bars are shown in purple. (C) Evolution over time of the position angle (PA); again, measured points and the corresponding error bars are shown in purple.}
\label{fig:c_orbits}
\end{figure*}

\begin{figure}
\centering
\includegraphics[width=\hsize]{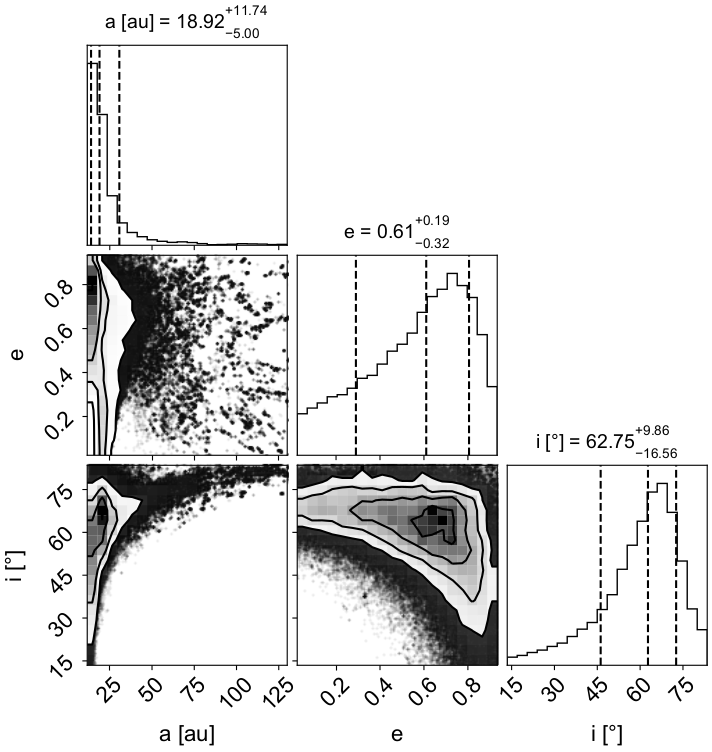}
\caption{Posterior distributions of orbital parameters for the probable CC0. Corner plot showing the posterior distributions from orbitize! for semi-major axis (a), eccentricity (e) and inclination (i) derived in the same way as in Figure~\ref{fig:b_param}.}
\label{fig:c_param}
\end{figure}

Starting from the measured separations and PA and their broadened uncertainties, orbital parameters were estimated separately for the probable companion CC0 and for the robust one $\mu^2$ Sco b using the orbitize! Python package\footnote{Available at \href{https://github.com/sblunt/orbitize}{https://github.com/sblunt/orbitize }.}, run with the recommended parameters for reliable convergence, for a total of $2 \cdot 10^6$ orbits \citep{2020AJ....159...89B}. The priors used for the MCMC were stellar mass and parallax, taken from Table~\ref{table:fitted_param}, as well as the star-planet separations and position angles, taken from Table~\ref{table:cc_info}. A subsample of suitable orbits, as well as the posterior distributions of semimajor axis, eccentricity, and inclination, are shown in Figure~\ref{fig:b_orbits}-\ref{fig:c_param}.

To understand whether the best-fit orbital configuration of the system can be dynamically stable, we refer to the Hill criterion. Let $M_*$ be the mass of the primary star, ($m_1$, $a_1$, $e_1$) and ($m_2$, $a_2$, $e_2$) the mass, semimajor axis and eccentricity of the inner and outer companion, respectively. Under the hypothesis that $m_1 \ll M_*$ and $m_2 \ll M_*$, a system is Hill stable, meaning that the two companions will avoid close approaches at all times, if:

\begin{equation}
    \alpha^{-3} \left ( \mu_1 + \frac{\mu_2}{\delta^2} \right ) (\mu_1 \gamma_1 + \mu_2 \gamma_2 \delta)^2 \geq 1+3^{4/3} \frac{\mu_1 \mu_2}{\alpha^{4/3}}
\end{equation}

\citet{1993Icar..106..247G}. Here $\mu_1=m_1/M_*$, $\mu_2=m_2/M_*$, $\alpha=\mu_1+\mu_2$, $\Delta=a_2-a_1$, $\delta=\sqrt{1+\Delta/a_1}$, $\gamma_1=\sqrt{1-e_1^2}$, $\gamma_2=\sqrt{1-e_2^2}$. 

Using nominal values for orbital parameters and masses, the orbits of b and CC0 are Hill stable; more accurately, taking the uncertainty on the orbital parameters into account, the system is Hill stable about $70 \%$ of the time. Likewise, the high nominal eccentricity values make it difficult to have additional stable orbits over a wide range of semimajor axes (from $\sim 5$ au up to $\sim 800$ au), though this cannot be excluded at present due to the large uncertainties still existing on the orbital parameters.

The final parameters of $\mu^2$ Sco b and the candidate CC0 derived throughout this Section are reported in Table~\ref{table:mu2_sco_companions}.

\begin{table*}
\caption{Absolute magnitudes, mass, projected separation, semimajor axis, eccentricity and inclination for the robust $\mu^2$ Sco b and the probable CC0. Here ($F_1$, $F_2$) = ($K_1$, $K_2$) for $\mu^2$ Sco b, ($J$, $H_2$) for CC0.}
\label{table:mu2_sco_companions}
\centering
\begin{tabular}{lcccccccc}
\hline\hline
& $F_1$ & $F_2$ & Mass & $q$ & Proj. sep & $a$ & $e$ & $i$\\
& mag & mag & $\mj$ & & au & au & & $^\circ$ \\
\hline 
$\mu^2$ Sco b & $9.86 \pm 0.33$ & $9.48 \pm 0.32$ & $14.4 \pm 0.8$ & 0.0015(1) & $290 \pm 10$ & $242.4^{+114.5}_{-52.1}$ & $0.56^{+0.27}_{-0.26}$ & $96.6^{+21.5}_{-20.5}$ \\
CC0 & $11.51 \pm 1.10$ & $9.69 \pm 0.38$ & $18.5 \pm 1.5$ & 0.0019(2) & $21 \pm 1$ & $18.9^{+11.7}_{-5.0}$ & $0.61^{+0.19}_{-0.32}$ & $62.8^{+9.9}_{-16.6}$ \\
\hline
\end{tabular}
\end{table*}

\section{Discussion}
\label{sec:discussion}

\begin{figure*}
  \begin{minipage}[c]{0.67\textwidth}
    \includegraphics[width=\textwidth]{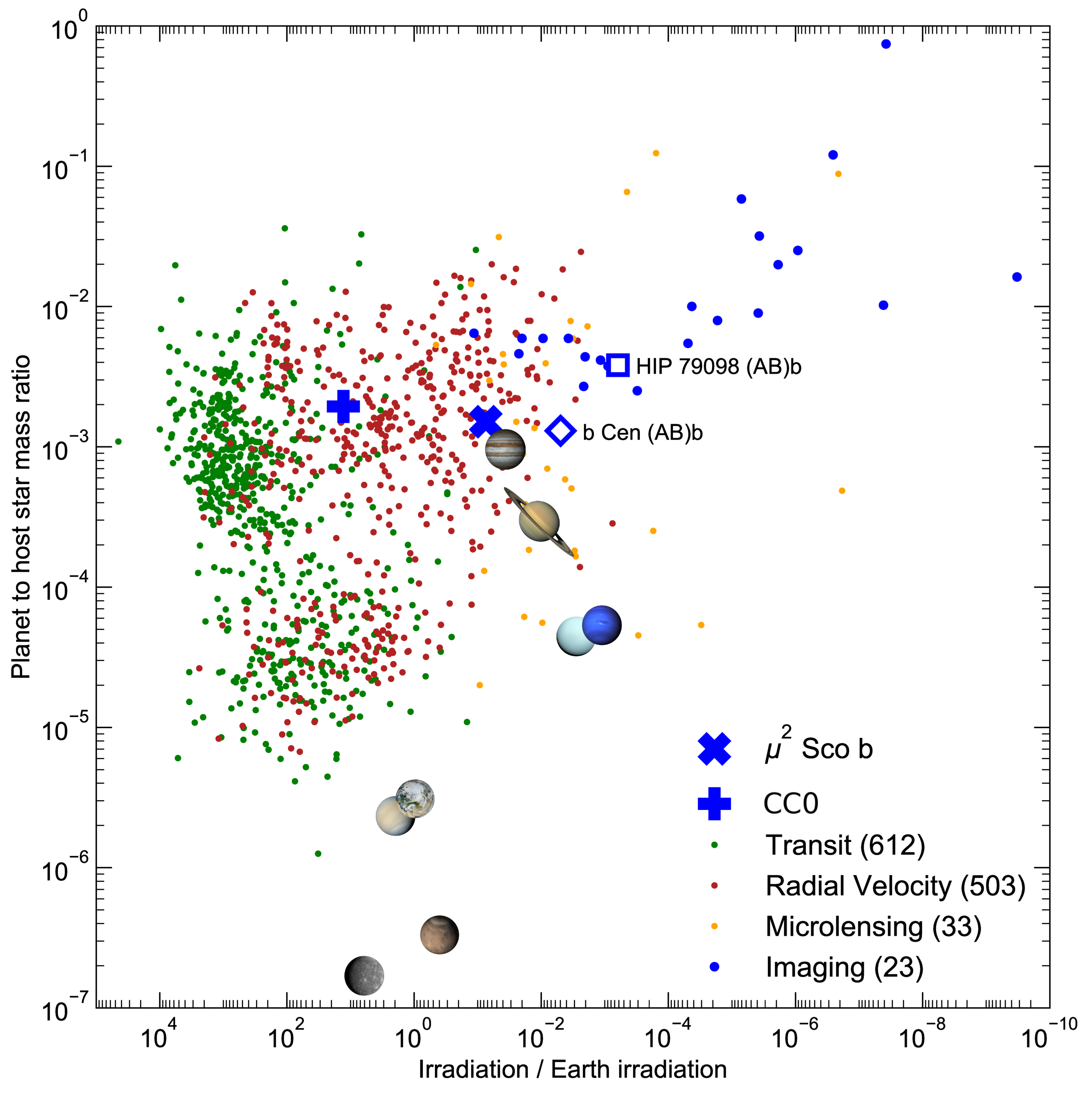}
  \end{minipage}\hfill
  \begin{minipage}[c]{0.3\textwidth}
    \caption{
        Mass ratio vs irradiation for known exoplanets. Only exoplanets whose stellar host mass is known to a precision of at least 30\% are shown. Each planet is labeled according to its detection method: transits in green, radial velocity in red, microlensing in orange and direct imaging in blue. BEAST discoveries are overplotted with larger blue symbols, and circular orbits with radius equal to the observed projected distance are assumed; Solar System planets (images from NASA) are also shown for reference. $\mu^2$ Sco b can be considered a Jupiter analog both in terms of irradiation and mass ratio (similar to the directly imaged 51 Eridani b \citep{2015Sci...350...64M}, obscured here by the icon for Jupiter) while the irradiation received by probable CC0 is similar to Mercury’s. Sources: NASA Exoplanet Archive (https://exoplanetarchive.ipac.caltech.edu/), the Extrasolar Planets Encyclopaedia (http://exoplanet.eu/).
    } \label{fig:irr_q_plot}
  \end{minipage}
\end{figure*}

The two objects, having $M_b=14.4 \pm 0.8 \mj$ and $M_c=18.5 \pm 1.5 \mj$, are just above the deuterium-burning limit ($M \sim 13 \mj$) that is classically used to mark the transition between planets and brown dwarfs. However, given the host mass, the mass ratios of these companions to the star are two of the lowest for objects discovered in imaging ($q_b=0.0015$, $q_c=0.0019$) and comparable to that of Jupiter to the Sun ($q=0.00095$). This qualifies them as planet-like from this point of view. Very interestingly, the mean irradiation that the two planet-like companions of $\mu^2$ Sco receive from their parent star is similar to those of two Solar System planets: while the outer one has a mean irradiation similar to that of Jupiter, the inner one should have an irradiation comparable to that of Mercury, that would make it, if confirmed, the most irradiated substellar companion discovered by direct imaging so far (Figure~\ref{fig:irr_q_plot}). The $\mu^2$ Sco system appears to be in many regards a scaled-up version of the Solar System.

Leaving aside the case to be confirmed of CC0, the robust discovery of a physical companion to $\mu^2$ Sco with a mass ratio similar to that of Jupiter to the Sun is not only noteworthy as an individual case, but acquires its greater significance when coupled to the recent discovery, within the same survey, of a similar object ($M=10.9 \pm 1.6 \mj$, $q=0.0013_{-0.0003}^{+0.0004}$) around the binary b Cen system. In that case the primary star has a mass of $\sim 6$~M$_\odot$\ and the overall mass is similar to that of $\mu^2$ Sco \citep{2021Natur.600..231J}. The possibility of capture of an object originally formed elsewhere, owing to the very low density of the Sco-Cen association to which both targets belong, was already unlikely for a single system\footnote{Even in the worst case of the relatively compact LS, the timescale for a close encounter between $\mu^2$ Sco and another star within 1000 au is $\sim 200$ Myr, using Eq. 1 from \citet{2007MNRAS.378.1207M} with $n=1 pc^{-3}$, $M_{tot}=10 M_\odot$ and $v=0.7$ km s$^{-1}$.}; the discovery of similar companions in two different systems firmly argues for in-situ formation, pointing toward a whole range of questions about how these objects came into being.

$\mu^2$ Sco b and b Cen b are added to the growing population of directly-imaged substellar companions \citep{2019ESS.....410002N,2021A&A...651A..72V}; as shown by Figure~\ref{fig:irr_q_plot}, this population is able to probe a unique niche of the parameter space, building upon the sensitivity of direct imaging to young massive companions at large separations from their stars.

Companions with $1 \mj \lesssim M \lesssim 40 \mj$ are more frequently found around stars more massive than the Sun in radial velocity studies \citep{2015A&A...574A.116R,2022arXiv220212800W}, and the trend has been independently confirmed by direct imaging surveys\footnote{Mass and separation ranges are highly varying across different direct imaging studies. See Table 3 from \citet{2021A&A...651A..72V} for a comprehensive comparison of the cited studies.} \citep{2016PASP..128j2001B,2019ESS.....410002N,2021A&A...651A..72V}. As mentioned in Section~\ref{sec:introduction}, radial velocity studies indicate a turnover in this trend at about 2~$M_\odot$; but the increasing separation of the snow line -- beyond which most giant planets are thought to form -- with stellar mass \citep{2008ApJ...673..502K}, coupled with the scarce sensitivity of RV to separations larger than a few au, do not allow to conclude whether a wide-orbit population around more massive stars exists. Direct imaging is therefore the only technique able to investigate the extreme limits of planet formation. Early results from the BEAST survey are showing that giant exoplanets or small brown dwarfs can form even around B stars, but the completion of the survey is necessary to insert these qualitative findings into a robust statistical framework; a full comparison of companion mass and separation distributions with those from other large direct imaging surveys like SHINE \citep{2017sf2a.conf..331C} or GPIES \citep{2019ESS.....410002N} is expected to quantify the relative contribution of different formation pathways, clarifying the role of stellar mass in shaping planetary systems.

We must ask ourselves at this point what we mean by "planet" and "planetary system", and if the $\mu^2$ Sco system should be seen as a planetary system or rather as a multiple stellar system composed of a massive star and one or two brown dwarfs. Just as the transition from substellar to stellar objects at 75–80~\mj is determined by the possibility to ignite hydrogen burning in their core, the transition from giant planets to brown dwarfs is often set to $\sim 13~\mj$, the so-called deuterium-burning limit (DBL). According to this definition, the $\mu^2$ Sco system should be considered a multiple stellar system, while b Cen b would be a (circumbinary) giant planet. The clear similarities between the two systems, though, highlight how a distinction that is based uniquely on a process happening in the companion core might not be adequate in every circumstance, and not necessarily related to different formation pathways.

On the other hand, a second distinction can be operated between objects being formed “like stars”, that is through turbulent core fragmentation, and objects being formed “like planets”, that is within a protoplanetary disk. This definition generally agrees with the previous one for solar-like stars: a correlation exists between the occurrence of giant planets with $M<4 \mj$ and stellar metallicity, hint of formation within a protoplanetary disk; more massive ($M\sim 10$~\mj) objects do not show this correlation \citep{2018ApJ...853...37S}, pointing toward a star-like formation process. However, when analyzing lower or higher stellar masses, the situation becomes more and more blurred: it is here that the mass ratio comes into play. On the one hand, a few known giant planet companions to very low-mass primary stars or brown dwarfs, likely outcome of turbulent fragmentation within the natal molecular cloud \citep[e.g.,][]{2020ApJ...905L..14F}, should be considered “star-like”. On the other hand, the two brown dwarfs companions ($M \sin i = 22~\mj$ and $25~\mj$) to the evolved 2.7 $M_\odot$ star $\nu$ Oph show a 6:1 mean motion resonance of their orbits, a clue of a formation within a protoplanetary disk, and should be labeled as “planet-like” \citep{2019A&A...624A..18Q}. This issue has been recently discussed within the IAU Commission F2, and led to a revised version of the definition of planets: although no explicit distinction based on the formation pathway (which is not easily related to physical properties) was set, an upper limit was established to the planet-to-star mass ratio of $q<0.04$ \citep[][]{2022arXiv220309520L} in addition to the DBL mass limit at $M<13 \mj$. While this revision is appropriate for systems around low-mass stars, a general observation is that, due to a scaling of disk mass with star mass \citep[see, e.g.,][]{2016ApJ...831..125P}, one could expect that more massive stars—originally surrounded by more massive protoplanetary disks— can form companions within a disk that are more massive than the deuterium-burning limit. We notice that the mass ratios $q$ of $\mu^2$ Sco b, the candidate CC0 and b Cen b are consistent with one another and -- within a factor two -- with that of Jupiter to the Sun (Figure~\ref{fig:irr_q_plot}); the mass ratios are smaller than those in the $\nu$ Oph system and the recently-adopted revised IAU definition of planet. As protoplanetary disks around massive stars can extend to thousands of au \citep{2016A&ARv..24....6B}, $\mu^2$ Sco b, the candidate CC0 and b Cen lie well within the primordial extent of the protoplanetary disk of their host stars. In this sense, $\mu^2$ Sco b and b Cen b should both be considered as planets.

A third way to assess whether $\mu^2$ Sco b and the candidate CC0 are to be considered “planets” is based on a purely empirical basis. Giant planets around less massive stars than $\mu^2$ Sco show a bottom-up mass distribution, that is, a larger occurrence of less massive planets; this is not true for stellar companions, which instead show a top-down distribution that favors larger values of $q$. This stellar companion population, by construction, encompasses every possible pathway for multiple star formation, ranging from turbulent core fragmentation, to fragmentation of pseudo-disks, to GI within a protoplanetary disk\footnote{The latter mechanism, according to the second distinction, would instead fall under the "planet" definition.}. One might wonder if objects like $\mu^2$ Sco b and b Cen b constitute the high-mass tail of the bottom-up planet-like population (PP), or rather the low-mass tail of the top-down star-like population (SP).

A tentative comparison between a SP and a PP -- based in turn on previous multiplicity and direct imaging studies, respectively -- can be made in a similar way as in \citet{2021Natur.600..231J}: for the SP, we adopt a log-normal separation distribution as in \citep{2014MNRAS.437.1216D} -- suited for A stars -- and a mass ratio distribution as in \citet{2013A&A...553A.124R}:
\begin{equation}
\frac{\partial^2 N}{\partial \log_{10}(a) \partial q} \propto \frac{(\log_{10}(a[au])-2.59)^2}{2\cdot 0.79^2} \cdot q^{0.25},
\end{equation}

\noindent normalizing it to the median frequency of $1-75~\mj$ companions at [5-300] AU for BA hosts taken from \citet{2021A&A...651A..72V}. For the PP, we adopt the parametric model from \citet{2021A&A...651A..72V} and \citet{meyer} with the set of parameters referring to BA stars:
\begin{equation}
\frac{\partial^2 N}{\partial \log_{10}(a) \partial q} \propto \frac{(\log_{10}(a[au])-0.79)^2}{2\cdot 0.77^2} \cdot q^{-1.31}.
\end{equation}

The expected number of companions around one star with $a \in [100, 1000]$ au, $q \in [0.0005, 0.0030]$ is $1.4 \cdot 10^{-4}$ for the SP scenario, $9.0 \cdot 10^{-3}$ for the PP scenario. Taking into account that we have observed 25 targets in the survey, the probability of finding at least one companion within these ranges is $3.4 \cdot 10^{-3}$ for the SP scenario and 0.20 for the PP scenario. If we further assume that -- as suggested for stars with $M<2.5 M_\odot$ \citep{2021A&A...651A..72V} -- the peak of the orbital distribution shifts to larger separations with stellar mass, and we employ irradiation levels as our scaling factor, the probability under the PP scenario rises to 0.79. In this last case, the probability of finding at least two companions -- as $\mu^2$ Sco b and b Cen b -- is fairly high too (0.44). We stress that the comparison is based on a naive extrapolation of the known frequencies of planetary and stellar companions from A stars to $\sim 9~M_\odot$ stars, and that a companion mass ratio distribution getting steeper at larger separations \citep{2014MNRAS.437.1216D,2017ApJS..230...15M} might mitigate --at least partially-- such a strong prevalence for the PP scenario.

From a formation standpoint, the bottom-up PP distribution is naturally associated with CA. CA assembles giant planets by building a solid core from dust present in the protoplanetary disk; as the core attains a critical mass ($\sim 10~M_\oplus$), it rapidly starts accreting hydrogen and helium to become a gas giant \citep{1996Icar..124...62P}. The shift of the orbital peak of the PP distribution would be a natural consequence of the wider distance of the snow line, and the increased reservoir of gas and dust in the protoplanetary disk would make it possible to form an object as massive as $\mu^2$ Sco b; in other words, the PP scenario would explain the similarity of $q$ ratio and irradiation with CA planets like Jupiter.

On the other hand, CA requires a few million years to operate, a timescale comparable to the disk lifetime around low-mass stars \citep{2009ApJ...705.1237G,2009ApJ...690.1539G}. The survival of a protoplanetary disk is mainly dictated by the strength of ultraviolet (FUV: $6~{\rm eV} <h \nu < 13.6~{\rm eV}$; EUV: $13.6~{\rm eV} < h\nu < 0.1~{\rm keV}$) and X-ray ($0.1~{\rm keV} < h\nu < 2~{\rm keV}$) irradiation from the central star. While disk lifetimes do not vary much for stellar masses in the range 0.3-3~M$_\odot$, more massive stars ($>7$~M$_\odot$) are expected to lose their disks rapidly (in $\sim 10^5$ years) due to extremely high EUV and FUV fields \citep{2009ApJ...690.1539G}. Indeed, only very feeble disk remnants have been observed around O stars, while disks around B stars survive for a few hundred thousand years, leaving not enough time for CA to operate \citep{2009ApJ...705.1237G}.

In this regard, it is interesting to qualitatively compare the disk survival timescales of the binary b Cen ($M_A \sim 5.5~M_\odot$, $M_B \sim 3.5~M_\odot$) and the single $\mu^2$ Sco ($M \sim 9.1~M_\odot$). While a naive comparison between b Cen A -- expected to emit nearly all the X and UV flux in the b Cen system and treated thus as a single star -- and $\mu^2$~Sco would yield a three times shorter disk lifetime around $\mu^2$~Sco \citep[see eq. 7 by ][]{2009ApJ...690.1539G}, the actual ratio should be much larger due to an initial disk mass for b Cen related to the total system mass, hence comparable to that of $\mu^2$ Sco. Combining Eq. (11) from \citet{2007prpl.conf..197C} for the photoevaporation outflow rate with the expected ionizing photon flux $\Phi$ for the three stars, the ratio\footnote{We interpolate between $\log{\Phi_{\rm EUV}}$ values for $M=3~M_\odot$ and $M=7~M_\odot$ \citep[see Table 2 by ][] {2009ApJ...690.1539G} and $\phi_i$ values for $M=16.8~M_\odot$, $M=25.6~M_\odot$ and $M=65~M_\odot$ \citep{1994ApJ...428..654H}, deriving the empirical relation  $\log{\Phi} = -5.9705 \log{M} + 21.553 \log{M} + 30.1$. The b Cen photoevaporation outflow rate is the sum of the individual contributions of b Cen A and b Cen B.} between the two disk survival timescales should be around $\sim 20$. While the impact of this on planet formation is difficult to be properly assessed, the presence of (at least) a companion with $q \sim 0.002$ around $\mu^2$ Sco looks much more challenging than that around b Cen in the framework of a CA scenario. Recent updates of the classical CA model, such as pebble accretion, have been indicated as a possible solution to the conundrum \citep{2012A&A...544A..32L}\footnote{Alternatively, the problem might be alleviated if $\mu^2$ Sco were actually formed by the merging of two nearly equal mass stars, which is possibly not an exotic case \citep{2014ApJ...782....7D}. On this respect, it is notable that $\mu^2$ Sco appears to be a slow rotator (see Appendix~\ref{sec:magnetic_field}); in fact it has been argued recently that post-mergers should appear as slow rotators \citet{2019Natur.574..211S, 2022NatAs.tmp...35W}. On the other hand, the same studies suggest that mergers might have strong magnetic fields, but there is no evidence for this in $\mu^2$ Sco (see Appendix~\ref{sec:magnetic_field}).}

On the other hand, giant planets may possibly form inside massive protoplanetary disks on a much shorter timescale of $\sim 10^4$~yr by means of GI. While this mechanism is not likely to be at the origin of most planets observed around solar-type stars, it might be considered for stars much more massive than the Sun; there is observational evidence of disk fragmentation around $10~M_\odot$ stars \citep{2020MNRAS.492.5041C,2021A&A...655A..84S}. GI preferentially produces massive planets in wide orbits, although rapid migration can force some of them to move much closer to the star \citep{2015ApJ...802...56M}. In this regard, it is interesting to notice that our orbital analysis shows a preference for large eccentricities for both the confirmed $\mu^2$ Sco b and the probable CC0. A recent analysis of 27 directly-imaged giant planets and brown dwarfs in the separation range 5-100 au around a wide range of stellar hosts ($0.2-2.8~M_\odot$) has hinted at an interesting eccentricity dichotomy between the two populations: while the former usually has low eccentricity values ($e=0.13^{+0.12}_{-0.08}$), the latter is characterized by a flat distribution over the range $0<e<1$ \citep{2020AJ....159...63B}. If this dichotomy continues at larger separations and stellar masses, it might favor an in-situ GI scenario for $\mu^2$ Sco b and CC0 (see Appendix~\ref{sec:formation_analysis}); alternatively, the large eccentricities might be simply a result of a strong dynamical evolution after formation, possibly causing the migration of $\mu^2$ Sco b and CC0 into their current orbits \citep[bringing a CA scenario back into the game; see, e.g.,][]{2019A&A...624A..20M}. We stress that any definitive conclusion on this point cannot be reached without a follow-up of the system in the next few years, aimed at both constraining the orbit of b and establishing if CC0 exists.

According to the parametric model described above, GI should be considered as one of the top-down processes underlying the SP scenario, preferentially producing massive BD and low-mass stars rather than giant planets \citep{2010ApJ...710.1375K}; indeed, extensive population synthesis simulations, that take into account migration and N-body interactions, find that 90-95\% of the surviving objects have masses above the DBL already around stars with $0.8-1.2 M_\odot$ \citep{2013MNRAS.432.3168F,2018MNRAS.474.5036F}. The corresponding typical mass ratio $q=0.01-0.1$ is one or two orders of magnitude larger than that of b Cen b and $\mu^2$ Sco b and CC0, implying an unusually low conversion efficiency of disk mass into companion mass for BEAST companions. Finally, although the completion of the survey is needed before drawing robust conclusions, the presence of already two or three companions near the deuterium-burning limit is not expected according to the mass distribution of objects generated by current GI models.

In this regard, recent 3D radiative magneto-hydrodynamic simulations have shown that, in a GI disk, the magnetic field acts to reduce the fragmentation scale by 1-2 orders of magnitude, possibly turning GI into a viable formation path for Jupiter-analogs around B stars \citep{2010Icar..207..509B,2021NatAs...5..440D}.

Irrespective of the formation scenario invoked, theoretical efforts toward the modeling of the exoplanet population have often been restricted to solar-like stellar hosts. First steps are being taken to extend toward lower masses (as a result of the attention drawn by M stars as potential hosts of habitable planets), but very little effort is going toward larger masses. This is mostly due to the lack of detections and the challenges of developing models without observational constraints. Upcoming BEAST observations will provide these crucial constraints, both by delivering a new population of companions and through nondetections, paving the way to a new suite of planet formation models around intermediate and massive stars.

The $\mu^2$ Sco system presented in this work appears remarkable in many regards: taken in isolation, it is the most massive star hosting a planetary system, pushing for the first time the frontier of exoplanetary studies well into the massive star regime; coupled with the b Cen system, it hints at an emerging companion population resembling a scaled-up version of the known giant planet population found around Sun-like stars; a population that, intriguingly, is not yet understood by the known mechanism of planet formation, urging for the development of new models. Follow-up of this unique system will be required to confirm the existence of CC0, and to better constrain the orbital configuration of the system, which in turn might give crucial clues on its formation history.

\section{Conclusions}
\label{sec:conclusions}

The BEAST program is devoted to the search of substellar -- possibly planetary -- companions at separations of tens to hundreds au from B stars ($M \gtrsim 2.4~M_\odot$) in the young (5-20~Myr old) nearby association Scorpius-Centaurus. The final aim of the program is to understand if the actual dearth of planets known around massive stars is due to an early dispersal of the protoplanetary disk or simply to a selection effect -- the planets being too far from the star to be discovered through RV searches. After the discovery of a planet companion to b Cen \citep{2021Natur.600..231J}, in this paper we have presented the intriguing case of $\mu^2$~Sco. After a careful re-evaluation of the kinematic properties of this star, we found it to belong to a small group of stars within Upper Centaurus-Lupus (UCL) that we have labeled Eastern Lower Scorpius (ELS). The star -- a slowly pulsating B-star $\beta$ Cep hybrid -- appears to be a slow rotator, and shows no evidence of a strong magnetic field; the combination of the two factors with a new distance estimate and suitable priors derived from the literature allowed us to determine the properties of the star. With an age of $20\pm 4$~Myr and a mass of $9.1\pm 0.3$~M$_\odot$, $\mu^2$ Sco appears to be a massive main-sequence star that will likely end its life as an electron-capture supernova.

Thanks to high-contrast imaging observations with SPHERE@VLT, we have detected a comoving source in IRDIS images with measured contrasts $\Delta K_1 \approx 11.7$ mag and $\Delta K_2 \approx 11.3$ mag. We examined the possibility that this object is a background interloper or small mass member of the Sco-Cen association unrelated to $\mu^2$ Sco and found both hypothesis to be highly improbable. We have therefore concluded that this object -- $\mu^2$ Sco b-- is a real physical substellar companion to the star. With a best-fit $M=14.4\pm0.8~\mj$, $\mu^2$ Sco b is the first "planet-like", in the sense defined in Section~\ref{sec:discussion}, companion to a supernova progenitor. 

A second $18.5\pm1.5~\mj$ substellar companion, CC0, is tentatively detected in IFS images, much closer to the star, but we highlight the need for follow-up observations to definitely confirm its existence. The object would be the most irradiated substellar companion ever discovered through direct imaging.

CC0 and $\mu^2$ Sco b have angular separations of $0.128\pm 0.002$~arcsec and $1.709\pm 0.005$~arcsec from the star that translate, at a distance $d=169\pm 6$~pc, into projected separations of $21\pm 1$~au and $290\pm 10$~au, respectively. The small astrometric displacements over the three years separating the two epochs (Figure~\ref{fig:astro_shifts}) are a consequence of their orbital motion around $\mu^2$ Sco: even though our observations are not enough to completely determine the orbits, we could constrain the set of orbital parameters that are consistent with observations. A preference for large inclinations and/or large eccentricities emerges; owing to the large b-c distance, the two-companion system should nonetheless be dynamically stable (Figure~\ref{fig:b_orbits}-\ref{fig:c_param}).

$\mu^2$ Sco b (and possibly CC0) are naturally added to b Cen b, a $10.9 \pm 1.6 \mj$ companion recently discovered around the binary b Cen system in the same BEAST survey \citep{2021Natur.600..231J}. While the survey is still ongoing and a sound statistical analysis is not yet possible, the discovery of already two (possibly three) companions with very small mass ratios -- similar to Jupiter's -- around the first 25 stars for which observations have been completed, dwelling at separations within the expected size of the protoplanetary disks, appears to be more compatible with a bottom-up "planet-like" formation scenario rather than a top-down "binary-like" scenario. The possible scaling of planetary system sizes with stellar mass would be in tension with the current definition of planets adopted by the IAU, which would classify the companion(s) of $\mu^2$ Sco as brown dwarfs and b Cen b as a giant planet.

The completion of the BEAST survey will allow a more definite answer to this issue, providing a census of companions around B stars over a wide range of masses. In addition, new observations of $\mu^2$ Sco with high-contrast imaging and near-IR interferometry are needed both to confirm the existence of CC0 and for a better characterization of the properties of $\mu^2$ Sco~b; coupled with asteroseismologic follow-up of the primary star, they will enable us to shed light on the fascinating history of this unique system.

\begin{acknowledgements}

    We are extremely grateful to the anonymous referee for the constructive comments, which significantly helped raise the quality of this paper.
    The observations have been acquired at the ESO VLT telescope (program 1101.C-0258).
    This work has made use of the SPHERE Data Center, jointly operated by OSUG/IPAG (Grenoble), PYTHEAS/LAM/CeSAM (Marseille), OCA/Lagrange (Nice), Observatoire de Lyon and Observatoire de Paris/LESIA (Paris). 
    This work has made use of data from the European Space Agency (ESA) mission Gaia (https://www.cosmos.esa.int/gaia), processed by the Gaia Data Processing and Analysis Consortium (DPAC, https://www.cosmos.esa.int/web/gaia/dpac/consortium). Funding for the DPAC has been provided by national institutions, in particular the institutions participating in the Gaia Multilateral Agreement. 
    This research has made use of the SIMBAD database, operated at CDS, Strasbourg, France. 
    This publication makes use of data products from the Two Micron All Sky Survey, which is a joint project of the University of Massachusetts and the Infrared Processing and Analysis Center/California Institute of Technology, funded by the National Aeronautics and Space Administration and the National Science Foundation. 
    Part of this research was carried out in part at the Jet Propulsion Laboratory, California Institute of Technology, under a contract with the National Aeronautics and Space Administration (80NM0018D0004).
    This work was supported by the Programme National de Planétologie (PNP) and ASHRA of CNRS/INSU, co-funded by CNES.
    VS, RG, DM, SD and VD acknowledge the support of PRIN-INAF 2019 Planetary Systems At Early Ages (PLATEA).
    Part of this work has been carried out within the framework of the National Centre of Competence in Research PlanetS supported by the Swiss National Science Foundation. NE, SPQ and LM acknowledge the financial support of the SNSF.
    GDM acknowledges the support of the Deutsche Forschungsgemeinschaft (DFG) priority program SPP 1992 “Exploring the Diversity of Extrasolar Planets” (MA 9185/1-1). GDM also acknowledges the support from the Swiss National Science Foundation under grant BSSGI0\_155816 “PlanetsInTime”.
    For the purpose of open access, the authors have applied a Creative Commons Attribution (CC BY) licence to any Author Accepted Manuscript version arising from this submission. 
\end{acknowledgements}

%
%

\bibliographystyle{aa}
\bibliography{biblio}

\begin{appendix}

\section{Optimization tests for stellar parameters}
\label{sec:optimization}
    
Our optimization tests for the parameters of $\mu^2$~Sco start from the prior distributions of $(M, R, \teff, \pi)$ described in Section~\ref{sec:star}. We create a synthetic sample of $10^7$ stars, each one described by a quadruplet $(M, R, \teff, \pi)$, where every parameter is randomly drawn from its prior distribution. The empirical table of intrinsic colors and temperatures of 5-30 Myr old stars by \citet{2013ApJS..208....9P} allows a straightforward conversion of bolometric luminosities into V magnitudes by means of calibrated bolometric corrections. Starting from the stellar luminosity expected for main sequence stars:
\begin{equation}
    \frac{L}{L_\odot}=\left ( \frac{R}{R_\odot} \right )^2 \cdot \left ( \frac{T}{T_\odot} \right )^4,
\end{equation}
we apply the interpolated bolometric correction as a function of $\teff$, the fixed extinction and the derived distance modulus to get synthetic apparent $V$\ magnitudes. A filter was then applied to select only the quadruplets that simultaneously satisfy the three conditions:
\begin{itemize}
\item $V \in [3.52,3.60]$, where the broadened error bar allows for some uncertainty in the bolometric correction itself;
\item $\log{g} \in [3.6,4.0]$;
\item a mass M such that $|M-M_L|/M<0.15$, where $M_L$ is the mass derived through a $L(M)$ relation valid for main sequence stars with $M>2~M_\odot$.
\end{itemize}
    
The $L(M)$\ relation (in solar units) is derived, for consistency, by fitting $L$\ as a function of $M$\ using the same tables:
\begin{equation}
\log{L}=0.47+3.36 \log{M}.
\end{equation}
    
Although the mean fractional error of the fitted points is about 4\%, we opted for a less stringent 15\% tolerance to account for possible deviations from the standard behavior of the underlying sample. We underline that this test does not allow to constrain the age of the star, since the tables are built by averaging over stars of different ages.
    
\begin{figure*}
\centering
\includegraphics[width=\hsize]{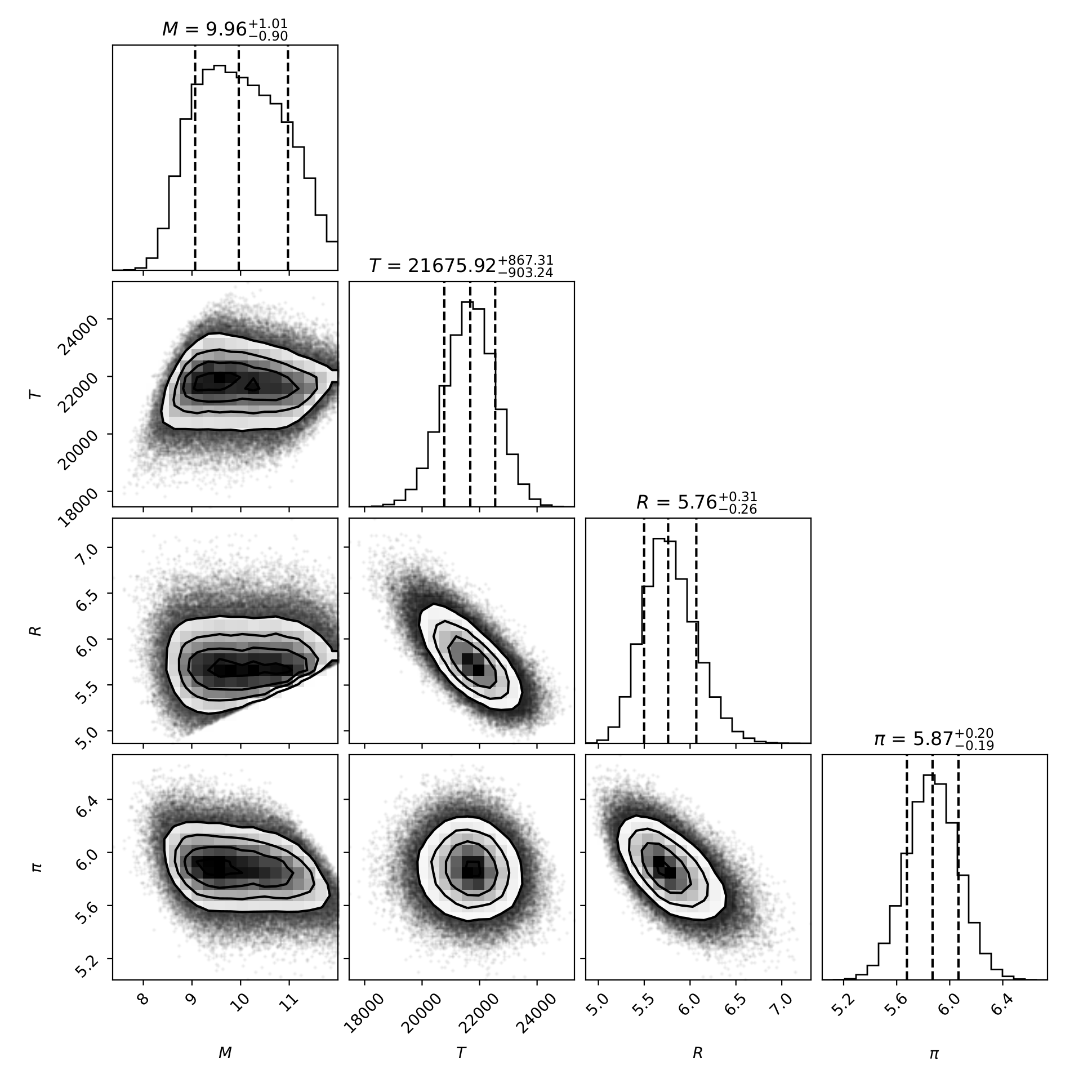}
\caption{Results of the first optimization test of stellar parameters. Corner plot showing the posterior distribution of the quadruplets $(M, R, \teff, \pi)$ consistent with the observational filter for the optimization test based on \citet{2013ApJS..208....9P}.}
\label{fig:star_param1}
\end{figure*}

The posterior distributions of $(M, R, \teff, \pi)$ are shown in Figure~\ref{fig:star_param1}. While the mean values of effective temperature and distance do not change significantly, a strong preference for a large value of radius appears ($R=5.8 \pm 0.3~R_\odot$), while the mass distribution shifts to very high values ($M=10.0_{-0.9}^{+1.0}~M_\odot$).
 
Considering that older stars on the main sequence are brighter, that is, age and mass are anti-correlated for MS stars of the same magnitudes, we expect our mass estimate to be corrected toward higher or lower values depending on the adopted age. In order to introduce the age into the discussion, we ran a similar but independent test that only relies on PARSEC isochrones \citep{2017ApJ...835...77M} of varying age and metallicity; drawing $10^7$\ masses and parallaxes as above, we selected the combinations that simultaneously satisfied:
\begin{itemize}
\item apparent Gaia magnitudes ($G, G_{BP}, G_{RP}$) within 0.1 mag from their observed values; absorption coefficients were taken from \citet{2019ApJ...877..116W};
\item a derived $\log (g) \in [3.6,4.0]$ as before;
\item $\teff \in [20900,22900]$ K.
\end{itemize}
    
\begin{figure*}
\centering
\includegraphics[width=\hsize]{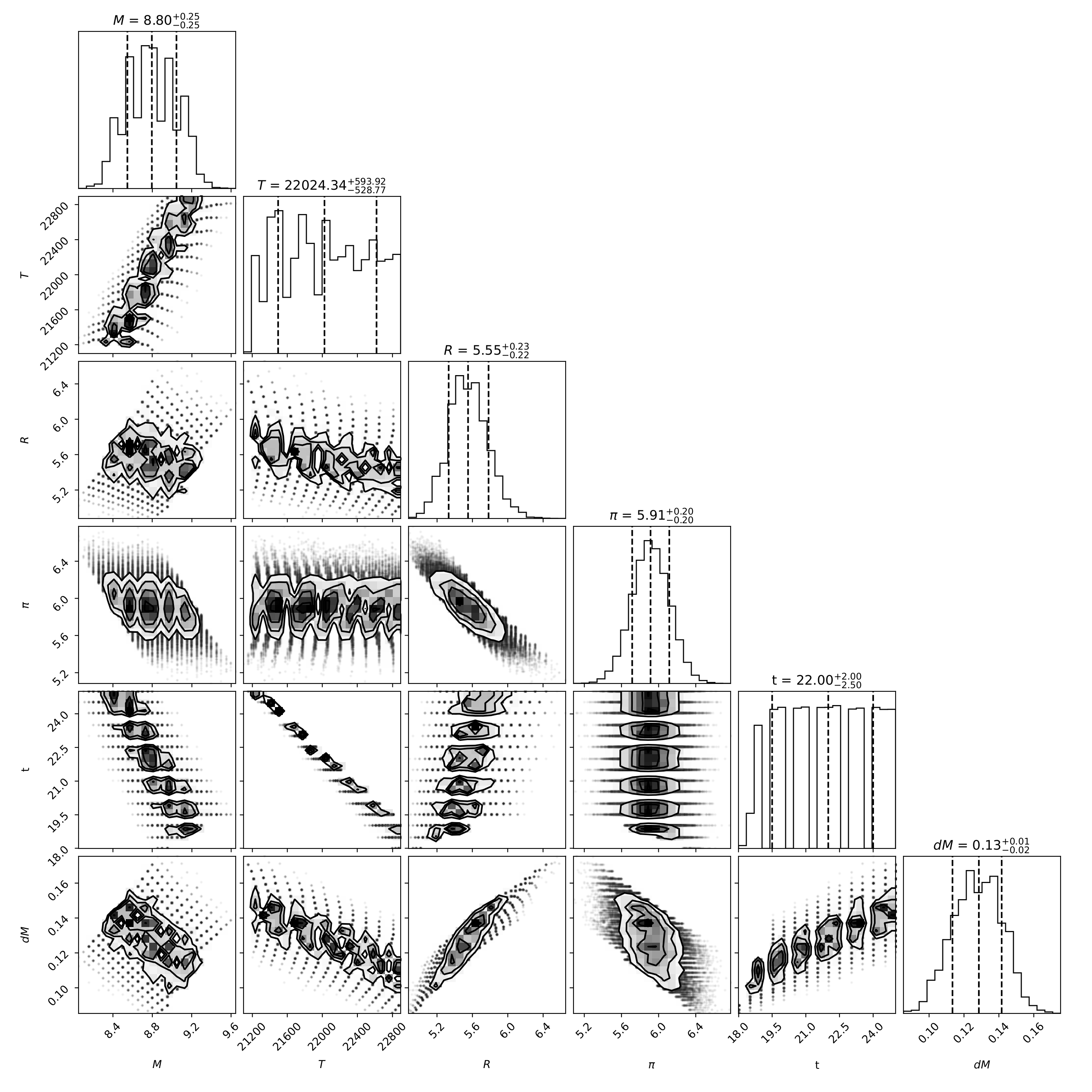}
\caption{Results of the second optimization test of stellar parameters (constant metallicity). Corner plot showing the posterior distribution of the doublets $(M, \pi)$ consistent with the observational filter for the optimization test based on PARSEC isochrones at fixed [Fe/H]=0.00. The corresponding $\teff$, $R$ and age distribution are shown too.}
\label{fig:star_param2}
\end{figure*}

\begin{figure*}
\centering
\includegraphics[width=\hsize]{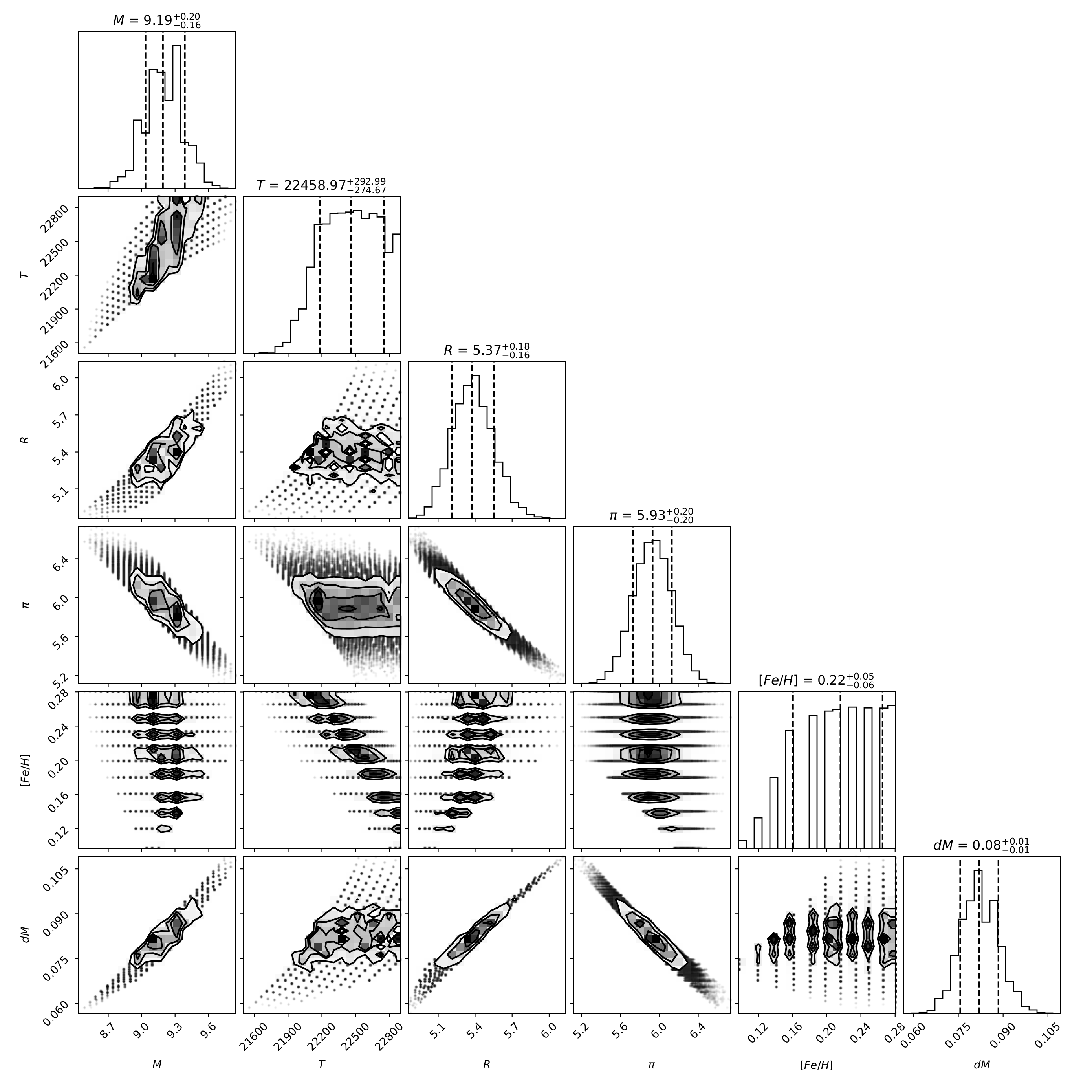}
\caption{Results of the second optimization test of stellar parameters (constant age). Corner plot showing the posterior distribution of the doublets $(M, \pi)$ consistent with the observational filter for the optimization test based on PARSEC isochrones at fixed $t=16$ Myr. The corresponding $\teff$, $R$ and [Fe/H] distribution are shown too.}
   \label{fig:star_param3}
   \end{figure*}

The test could never be passed using the nominal age ($t=16$~Myr) and metallicity ([Fe/H]=0.00), but increasing at least one of the two constraints ($t \in [16,25]$ Myr or [Fe/H] $\in [0.10,0.20]$) a family of solutions appeared (Figure~\ref{fig:star_param2}-\ref{fig:star_param3}); the posterior mass distribution is consistent with the result of the first test ($M=8.8 \pm 0.3~M_\odot$ for the run at constant metallicity, $M=9.2 \pm 0.2~M_\odot$ for the run at constant age; we may combine the two to get $M=9.0 \pm 0.3~M_\odot$). We average the two mass determinations from the independent methods to a derive a final $M=9.1 \pm 0.3~M_\odot$. In the same way, we derive $R=5.6 \pm 0.2~R_\odot$; the median values of the posteriors of the first test will be used as best-fit estimates for $\teff$ and $\pi$. Finally, while we are not able to solve the age-metallicity degeneracy, we encapsulate the general tendency for a larger age in a revised age estimate $t=20 \pm 4$ Myr.

\section{Possible binarity of \texorpdfstring{$\mu^2$}{mu2} Sco}
\label{sec:binarity}

The astrometric solution of $\mu^2$ Sco in the literature appears highly problematic (Table~\ref{table:astrom_data}). As shown by \citet{2019A&A...623A..72K}, a significant deviation of proper motion components from the long-term motion reconstructed by cross-matching Hipparcos and Gaia can be suggestive of the gravitational effects from an unseen stellar companion. Given the importance of the system architecture for a correct characterization and interpretation of our results, we tried to put constraints on possible unresolved stellar companions.

\begin{table}[!htbp]
\caption{Astrometric solutions for $\mu^2$ Sco in the recent literature. The values adopted throughout this paper combine proper motions by \citet{2019A&A...623A..72K} and the kinematic parallax based on membership to ELS. References: (1): Hipparcos (1997 reduction); (2): Hipparcos (2007 reduction); (3) Gaia DR2; (4) Gaia EDR3; (5) Kervella et al. (2019).}
\label{table:astrom_data}
\centering
\begin{tabular}{cccc}
\hline\hline
Source & $\pi$ & $\mu_\alpha^*$ & $\mu_\delta$ \\
& mas & mas yr$^{-1}$ & mas yr$^{-1}$ \\
\hline          
(1) & $6.31 \pm 0.86$ & $-12.92 \pm 0.66$ & $-23.80 \pm 0.61$ \\
(2) & $6.88 \pm 0.12$ & $-11.09 \pm 0.13$ & $-23.32 \pm 0.11$ \\
(3) & $7.92 \pm 0.55$ & $-9.98 \pm 0.96$ & $-19.87 \pm 0.78$ \\
(4) & $5.66 \pm 0.28$ & $-12.11 \pm 0.30$ & $-22.57 \pm 0.27$ \\
(5) & \textemdash & $-11.77 \pm 0.02$ & $-23.11 \pm 0.02$ \\
\hline          
adopted & $5.9 \pm 0.2$ & $-11.77 \pm 0.02$ & $-23.11 \pm 0.02$ \\
\hline          
\end{tabular}
\end{table}

Although \citet{2019A&A...623A..72K} reports a significant (S/N=4.1) proper motion anomaly (PMA) at Hipparcos era, the PMA was not significant at Gaia DR2 era. Thus, we checked the consistency of the astrometric solution from Hipparcos, finding that the 2007 reduction yields a PMA along right ascension with an opposite sign and similar magnitude (about 0.7 mas yr$^{-1}$) compared to the 1997 reduction based on the same raw data \citep{1997A&A...323L..49P}. Therefore, the tabulated PMA of Hipparcos cannot be trusted for this star. We use the nonsignificant Gaia DR2 PMA as a separation-dependent upper limit on the stellar companion mass.

$\mu^2$ Sco is neither a visual binary \citep{2021A&A...649A...1G} nor an interferometric binary \citep{2013MNRAS.436.1694R}; additionally, it does not appear as an X-ray source in ROSAT \citep{1996A&AS..118..481B}. The upper limit on the X-ray luminosity, given in units of $L_X=\log{f_x}$ [erg s$^{-1}$], is $L_X=29.82$. Since the estimate assumes a distance $d \sim 206$ pc, recalibration with the distance adopted in this work yields $L_X=29.65$. Such an emission is below the predicted plateau of a K5 star \citep{2014A&A...565L...1P}, posing an upper limit, independent of separation, of $M_2<0.7$~M$_\odot$.
 
We then analyzed HARPS \citep{2003Msngr.114...20M} and FEROS \citep{1999Msngr..95....8K} radial velocity data, spanning about 10 years retrieved from the ESO Science Archive\footnote{Data available at \href{http://archive.eso.org/scienceportal/home}{http://archive.eso.org/scienceportal/home, Programs 69.D-0677, 091.C-0713, 187.D-0917}.}: after extraction of RV from 37 He I and atomic lines, we verified that the rms of the observations (0.21 km s$^{-1}$) is smaller than individual uncertainties ($\sim 0.3$ km s$^{-1}$). Spectra taken at short temporal separation can push the sensitivity further: the four HARPS observations provide a small scatter of 74 m/s, and the two FEROS observations acquired on JD=56523.66 only differ by 90 m/s. The overall shallow trend of $\sim 50$ m s$^{-1}$ yr$^{-1}$ can be used to derive an additional upper mass limit: we used the Exo-DMC code \citep{2020ascl.soft10008B} under default assumptions to derive a 95\% confidence interval for this mass limit.
 
\begin{figure}
\centering
\includegraphics[width=\hsize]{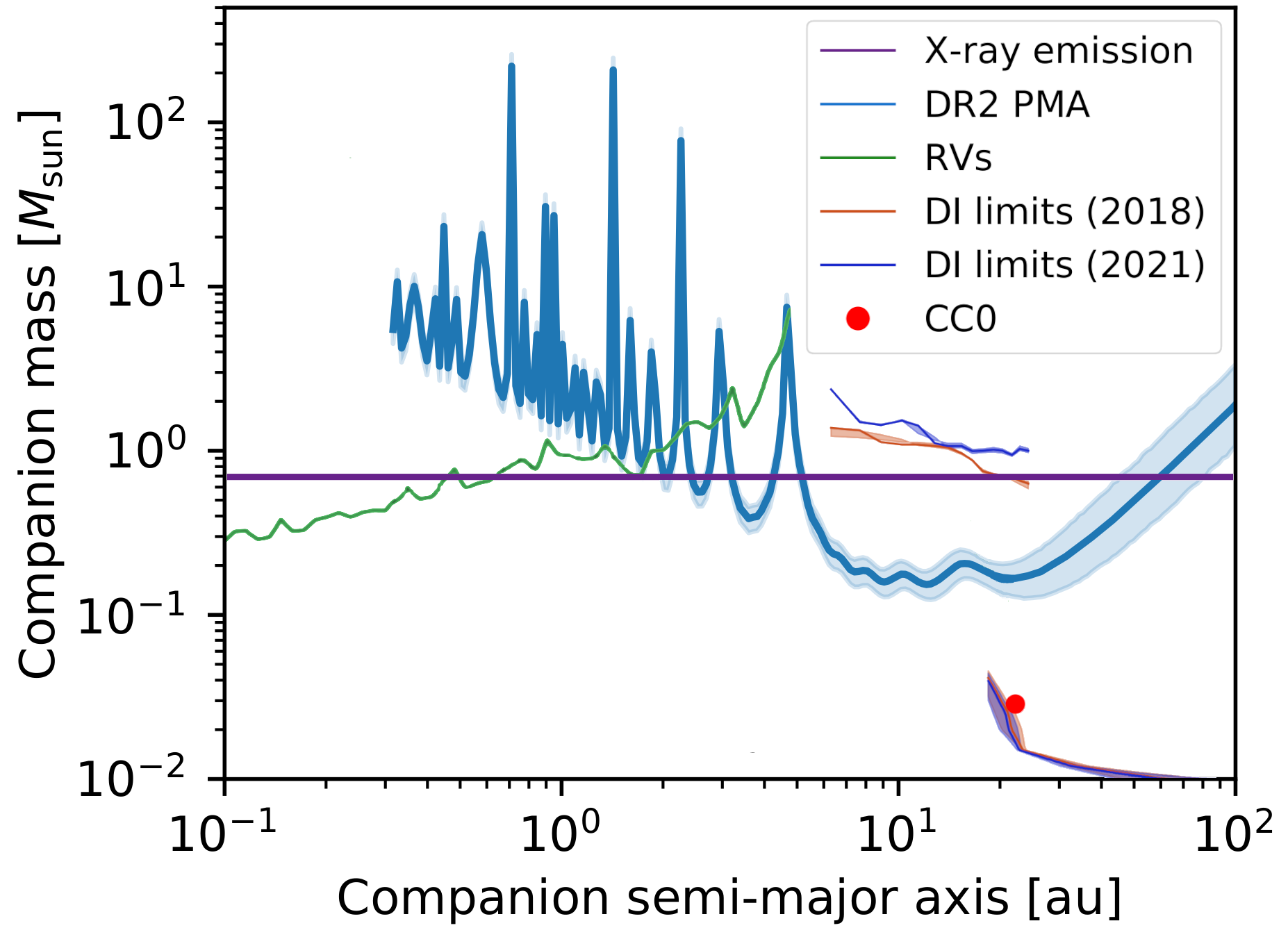}
   \caption{Limits on the mass of a possible unresolved stellar companion coming from several techniques. A $M>0.3-0.5~M_\odot$ companion is excluded at sub-au separations by RVs (95\% confidence interval), a $M>0.7~M_\odot$ is excluded at 1-5 au by the lack of X-ray detection. Astrometry (PMA) and coronagraphic observations place even stronger limits at $d>5$ au. At $\sim 290$ au, $\mu^2$ Sco b lies outside the x range.}
\label{fig:binary_mass_limits}
\end{figure}

Looking at Figure~\ref{fig:binary_mass_limits}, we can reasonably exclude the presence of a close stellar companion with $q>0.08$; additionally, we do not have any evidence to support the existence of a smaller stellar companion. We notice that the parallax error in the latest Gaia release is in line with that expected in Gaia EDR3 for a source with $G=3.5$ \citep{2021A&A...649A...2L}: the high errors associated with Gaia's astrometric solution are likely a mere consequence of the extreme brightness of the star.

\section{Stellar magnetic fields and rotation}
\label{sec:magnetic_field}

As the presence of a strong magnetic field can hinder a reliable comparison with model isochrones (see Section 2), we analyzed three high-resolution spectra of $\mu^2$ Sco obtained as part of the Magnetism in Massive Stars (MiMeS) project \citep{2016MNRAS.456....2W}: 
two taken by the spectropolarimeter ESPaDOnS \citep{2006ASPC..358..362D} in 2014 and 2015, and one taken by spectropolarimeter HARPSpol \citep{2011Msngr.143....7P} in 2011. To begin with, comparison of the intensity spectrum (Stokes I) with a non-LTE TLUSTY synthetic model \citep{2011ascl.soft09021H} yields values for the stellar parameters that are fully consistent with those derived through our optimization tests. We analyzed both Stokes I and Stokes V (circular polarization) spectra applying the Least-Squares Deconvolution technique (LSD) to perform a sort of weighted mean in all the spectral lines \citep{1997MNRAS.291..658D}. This method provides mean photospheric Stokes I and V profiles with a S/N much better than in the individual lines, and allows therefore to put stringent constraints on the surface magnetic field. 

The weights used in LSD are the predicted central depth of the intensity lines, the wavelength and the Landé factor. The mask, that is the list of predicted lines, has been obtained by using the VALD atomic line database\footnote{Available at \href{http://vald.astro.uu.se/}{http://vald.astro.uu.se/}.}, removing only some usually very strong lines (Balmer and He I) affected by non-LTE effects. 

\begin{figure}
\centering
\includegraphics[width=\hsize]{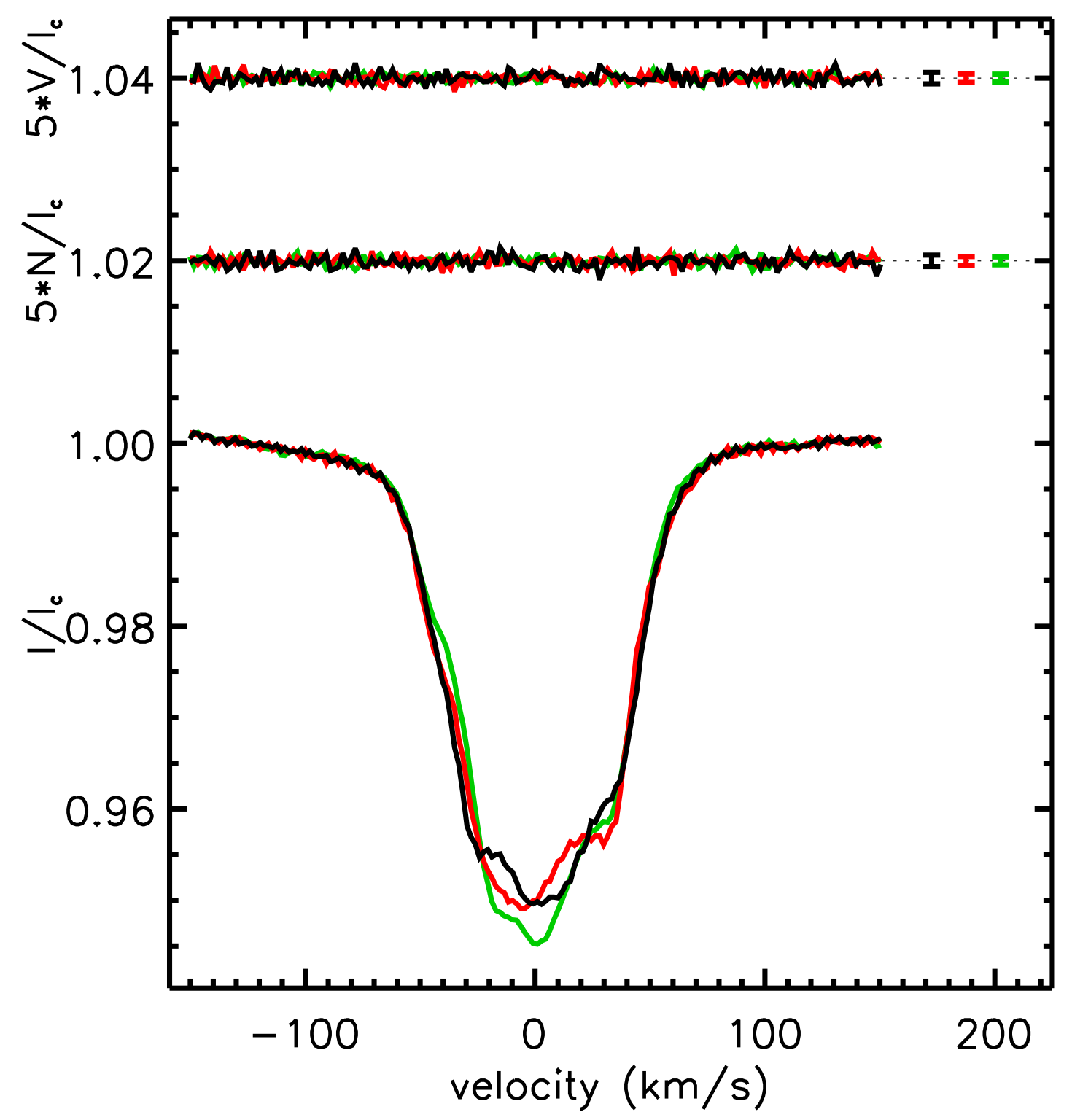}
   \caption{Resulting LSD profiles for the three $\mu^2$ Sco spectra (black: ESPaDOnS 2014; red: ESPaDOnS 2015; green: HARPSpol 2011)}. Top profile: Stokes V. Middle profile: null profile. Bottom profile: mean Stokes I profile. For each observation, the typical errorbars of individual spectral points are shown on the right. A vertical offset was applied to the top and the middle curve for better visibility.
\label{fig:magnetic_field}
\end{figure}

The resulting LSD profiles are shown in Figure~\ref{fig:magnetic_field}. The bottom profile shows the mean Stokes I profile, typical of a photospheric profile of a massive star, with broad wings indicating a significant macroturbulent velocity. The profile is clearly disturbed by the $\beta$ Cephei pulsations. The top profile is Stokes V, and the middle one is the null profile N, computed in a way that allows us to check that no spurious polarization is present in our data. All the curves are normalized to the mean continuum intensity, $I_c$.

The Stokes V profile is totally flat, indicating that there is no Zeeman detection in the data. All the measurements of the line-of-sight component of the magnetic field averaged over the surface of the star, called the longitudinal magnetic field Bl, are consistent with 0 G with an uncertainty of $\sim 15 G$. Using a Monte Carlo simulation \citep{2016A&A...589A..47A}, we estimate that if a dipole field at the pole with $B \geq 170$ G exists, we would have had a 90\% chance to have detected it. As the typical magnetic fields of massive stars are between a few hundreds of Gauss to few kG, $\mu^2$ Sco is not likely to host a strong fossil field.

With respect to rotational velocity, the low observed $v \sin{i} = 52$ km s$^{-1}$ implies a geometric probability of having $v>100$ km s$^{-1}$ of about 13\%. By comparison, the median $v \sin{i}$ for B stars belonging to $\sim 20$ Myr regions has been estimated as $\sim 125$ km s$^{-1}$ \citep{2007AJ....133.1092W}. $\mu^2$ Sco appears therefore to be a slow rotator.
   
\section{Formation and dynamical analysis of the system under a GI scenario} 
\label{sec:formation_analysis}

Although the large eccentricities of both the robust $\mu^2$ Sco b and the probable CC0 can be suggestive of a strong dynamical evolution after their birth, it is nevertheless instructive to see whether an in-situ formation at the position of CC0 is possible according to current formation models. With a best-fit semimajor axis of about 20 au, the mean irradiation level of CC0 is comparable to that of Mercury, preventing the presence of ice grains which are fundamental in a CA scenario \citep{2012A&A...541A..97M}. Whether the same argument applies to GI is not clear, crucially depending on the values of Toomre’s Q parameter and the cooling time, which determine if the protoplanetary disk can fragment or not at this position. Fragmentation occurs if Toomre’s Q parameter, defined as:
\begin{equation}
Q = \frac{\Omega^* c_s}{\pi G \Sigma},
\end{equation}
is less than unity. Here $\Omega^*$ is the epicyclic frequency, $c_s$ is the sound speed, $G$ is the gravitational constant and $\Sigma$ is the surface density. Since the radial dependencies of $\Omega^*$ and $\Sigma$ are similar ($\propto r^{-3/2}$), the radial variation of $Q$ depends on that of $c_s$, and the treatment can be simplified. Recalling that a full modeling of GI around stars as massive as $\mu^2$ Sco is still lacking, we might start from simple scaling laws from the results obtained for solar hosts. The relevant proportionalities in Q are given by: $\Omega^* \propto M_{star}^{1/2}$,  $c_s \propto T_{gas}^{1/2}$, $\Sigma \propto M_{disk}$. As regards the disk mass, we conservatively assume a linear proportionality between $M_{disk}$ and $M_{star}$. For the gas temperature, we might start from our best-fit estimates for stellar radius and $\teff$ to derive the effect due to stellar irradiation at 20 au (Eq. 4, 96) that is $T_{irr}=690\pm 50$ K, supposing a flared disk in vertical hydrostatic equilibrium. The onset of fragmentation for $0.1 M_\odot$ disks around $1 M_\odot$ stars in theoretical simulations occurs for $T \lesssim 50$ K \citep{2014prpl.conf..643H}, so we may write:
\begin{equation}
Q_c = \frac{\sqrt{\frac{M_{star}}{1 M_\odot}}\sqrt{\frac{T_{irr}}{50 K}}}{\left ( \frac{M_{disk}}{0.1 M_\odot} \right )} \sim 1.23.
\end{equation}
Since fragmentation happens if $Q_c<1$, a formation from GI would not be possible at the location of CC0. However, if $M_{disk}/M_{star}$ increases by more than 20\% than expected from the linear proportionality in the mass range of B stars, as assumed above, in situ fragmentation at the position of CC0 becomes possible. Indeed, indications exist that -- at least for $M<2 M_\odot$ -- $M_{disk} \propto M^q$  with $q=1.3-2.0$ \citep{2016ApJ...831..125P}: assuming that $q>1.1$ for B stars, the value of $Q_c$ would become lower than 1. Also, a warmer disk reduces the cooling time, favoring the onset of fragmentation at fixed $Q$. On the other hand, given the much lower irradiation temperature at the distance of $\mu^2$ Sco b, $\mu^2$ Sco b could have easily formed in situ via GI: not only the Toomre parameter is likely much smaller than unity in the approximations considered above; but also, more importantly, the conditions in protoplanetary disks at $R > 50$ au are most favorable to fragmentation since gas, even in a massive disk, is expected to be optically thin, thus leading to short cooling timescales \citep{2010Icar..207..509B}.

An interesting consequence of this scenario would be that disks around stars similar to $\mu^2$ Sco might be unstable at some phase of their evolution over a wide range of separations. However, the formation of planet-like objects such as the companions of b Cen and $\mu^2$ Sco rather than more massive (even stellar) objects would imply an extremely low efficiency of companion accretion, with the rest either accumulated on other objects (the star itself or a stellar companion) or lost from the system.

\onecolumn

\section{Companion candidates}
\label{sec:cc_tables}
\begin{table*}[!htbp]
\caption{Astrometric and photometric properties of the CCs. The contrasts ($F_1$, $F_2$) = ($K_1$, $K_2$) for IRDIS, and ($J$, $H_2$) for IFS.}
\label{table:cc_info}
\centering
\scalebox{0.9}{
\[
\begin{array}{cc|cccc|cccc}
\hline \hline
\multicolumn{2}{c}{} & \multicolumn{4}{c}{\text{First epoch}} & \multicolumn{4}{c}{\text{Second epoch}} \\
\hline
\multicolumn{10}{c}{\text{Comoving companions}} \\
\hline
\text{ID} & \text{Instr.} & d \text{ (mas)} & \text{PA} (^\circ) & F_1 \text{ (mag)} & F_2 \text{ (mag)} & d \text{ (mas)} & \text{PA} (^\circ) & F_1 \text{ (mag)} & F_2 \text{ (mag)} \\
\hline        
0 & \text{IFS} & 127 \pm 2 & 99.7 \pm 1.1 & 13.50 \pm 1.05 & 11.68 \pm 0.29 & 124.7 \pm 2.8 & 111.4 \pm 1.5 & \text{\textemdash} & \text{\textemdash} \\
2 & \text{IRDIS} & 1704 \pm 2 & 234.52 \pm 0.05 & 11.72 \pm 0.07 & 11.34 \pm 0.06  & 1714 \pm 1 & 234.47 \pm 0.05 & 11.73 \pm 0.06 & 11.35 \pm 0.05 \\
\hline        
\multicolumn{10}{c}{\text{Background sources}} \\
\hline
\multirow{ 2}{*}{1} & \text{IFS} & 722 \pm 6 & 132.7 \pm 0.4 & \text{\textemdash} & \text{\textemdash} & 703 \pm 3 & 127.2 \pm 0.2 & \text{\textemdash} & 15.64 \pm 0.20 \\
& \text{IRDIS} & \text{\textemdash} & \text{\textemdash} & \text{\textemdash} & \text{\textemdash} & 715 \pm 6 & 127.08 \pm 0.55 & 15.25 \pm 0.11 & 14.99 \pm 0.37 \\
3 & \text{IRDIS} & 2214 \pm 11 & 24.58 \pm 0.26 & 16.13 \pm 0.16 & 15.67 \pm 0.24 & 2282 \pm 6 & 24.73 \pm 0.16 & 16.00 \pm 0.13 & 15.57 \pm 0.27 \\
4 & \text{IRDIS} & 2478 \pm 3 & 236.55 \pm 0.06 & 13.55 \pm 0.07 & 13.43 \pm 0.07 & 2424 \pm 1 & 237.53 \pm 0.06 & 13.51 \pm 0.06 & 13.46 \pm 0.06 \\
5 & \text{IRDIS} & 2481 \pm 6 & 29.84 \pm 0.15 &  15.50 \pm 0.13 & 15.14 \pm 0.18 & 2527 \pm 5 & 29.63 \pm 0.13 & 15.81 \pm 0.10 & 15.52 \pm 0.18 \\
6 & \text{IRDIS} & 2633 \pm 9 & 259.42 \pm 0.15 & 16.07 \pm 0.18 & \text{\textemdash} & 2594 \pm 9 & 260.38 \pm 0.17 & 16.23 \pm 0.14 & 
\text{\textemdash} \\
9 & \text{IRDIS} & 2999 \pm 6 & 295.39 \pm 0.11 & 15.29 \pm 0.12 & 15.43 \pm 0.35 & 3007 \pm 3 & 296.40 \pm 0.09 & 15.25 \pm 0.08 & 15.11 \pm 0.14 \\
12 & \text{IRDIS} & 3491 \pm 6 & 27.84 \pm 0.09 & 15.07 \pm 0.09 & 14.88 \pm 0.30 & 3566 \pm 3 & 27.72 \pm 0.07 & 14.91 \pm 0.07 & 14.77 \pm 0.09 \\
13 & \text{IRDIS} & 3649 \pm 8 & 257.86 \pm 0.54 & \text{\textemdash} & 16.41 \pm 0.38 & 3588 \pm 2 & 258.13 \pm 0.17 & 16.72 \pm 0.17 & 16.33 \pm 0.40 \\
14 & \text{IRDIS} & 3703 \pm 6 & 10.66 \pm 0.08 & 15.40 \pm 0.09 & 15.45 \pm 0.29 & 3775 \pm 4 & 10.77 \pm 0.07 & 15.36 \pm 0.08 & 15.06 \pm 0.12 \\
15 & \text{IRDIS} & 3731 \pm 6 & 20.78 \pm 0.09 & 15.59 \pm 0.15 & 15.36 \pm 0.21 & 3780 \pm 4 & 20.89 \pm 0.08 & 15.58 \pm 0.08 & 15.57 \pm 0.18 \\
16 & \text{IRDIS} & 3748 \pm 1 & 50.70 \pm 0.60 & \text{\textemdash} & 16.30 \pm 0.45 & 3798 \pm 8 & 50.35 \pm 0.16 & 16.41 \pm 0.16 & 16.25 \pm 0.29 \\
17 & \text{IRDIS} & 3847 \pm 7 & 142.84 \pm 0.14 & 15.76 \pm 0.12 & 15.62 \pm 0.28 & 3812 \pm 5 & 141.99 \pm 0.12 & 15.67 \pm 0.08 & 15.78 \pm 0.40 \\
18 & \text{IRDIS} & 4142 \pm 4 & 136.94 \pm 0.05 & 12.02 \pm 0.07 & 11.95 \pm 0.06 & 4127 \pm 1 & 135.90 \pm 0.05 & 11.99 \pm 0.06 & 11.90 \pm 0.05 \\
19 & \text{IRDIS} & 4424 \pm 6 & 118.12 \pm 0.07 & \text{\textemdash} & 12.86 \pm 0.07 & 4422 \pm 2 & 117.24 \pm 0.05 & 12.90 \pm 0.06 & 12.81 \pm 0.05 \\21 & \text{IRDIS} & 4842 \pm 7 & 149.25 \pm 0.11 & 15.65 \pm 0.12 & 15.65 \pm 0.70 & 4812 \pm 4 & 148.66 \pm 0.11 & 15.71 \pm 0.09 & 15.69 \pm 0.22 \\
22 & \text{IRDIS} & 4974 \pm 11 & 132.12 \pm 0.16 & 16.51 \pm 0.28 & \text{\textemdash} & 4965 \pm 8 & 131.32 \pm 0.14 & 16.54 \pm 0.15 & 16.68 \pm 0.70 \\
23 & \text{IRDIS} & 4997 \pm 6 & 192.24 \pm 0.13 & 14.91 \pm 0.08 & 14.64 \pm 0.18 & 4944 \pm 3 & 192.05 \pm 0.13 & 14.88 \pm 0.07 & 14.81 \pm 0.09 \\
24 & \text{IRDIS} & 5043 \pm 11 & 178.95 \pm 5.93 & 16.58 \pm 0.22 & 15.97 \pm 0.87 & 4986 \pm 9 & 178.42 \pm 3.47 & 16.60 \pm 0.13 & \text{\textemdash} \\
26 & \text{IRDIS} & 5166 \pm 5 & 99.50 \pm 0.05 & 14.29 \pm 0.07 & 14.20 \pm 0.08 & 5192 \pm 2 & 98.69 \pm 0.05 & 14.25 \pm 0.06 & 14.23 \pm 0.07 \\
27 & \text{IRDIS} & 5185 \pm 11 & 117.99 \pm 0.13 & 16.34 \pm 0.25 & 16.18 \pm 0.35 & 5189 \pm 9 & 117.21 \pm 0.11 & 16.45 \pm 0.15 & 15.85 \pm 0.42 \\
28 & \text{IRDIS} & 5226 \pm 18 & 284.20 \pm 0.20 & 17.27 \pm 0.30 & 16.66 \pm 0.40 & 5222 \pm 11 & 284.78 \pm 0.13 & 16.76 \pm 0.18 & 
\text{\textemdash} \\
31 & \text{IRDIS} & 5294 \pm 11 & 295.91 \pm 0.10 & 16.16 \pm 0.13 & \text{\textemdash} & 5298 \pm 6 & 296.68 \pm 0.08 & 16.22 \pm 0.11 & 15.86 \pm 0.32 \\
32 & \text{IRDIS} & 5366 \pm 6 & 103.59 \pm 0.05 & 14.34 \pm 0.07 & 14.22 \pm 0.08 & 5382 \pm 3 & 102.89 \pm 0.05 & 14.29 \pm 0.06 & 14.20 \pm 0.07 \\ 
33 & \text{IRDIS} & 5418 \pm 7 & 36.73 \pm 0.07 & 14.96 \pm 0.10 & 14.86 \pm 0.13 & 5491 \pm 4 & 36.64 \pm 0.07 & 14.89 \pm 0.08 & 14.80 \pm 0.11 \\
36 & \text{IRDIS} & 5651 \pm 9 & 297.64 \pm 0.09 & \text{\textemdash} & 14.31 \pm 0.12 & 5644 \pm 2 & 298.22 \pm 0.05 & 14.33 \pm 0.06 & 14.17 \pm 0.08 \\
37 & \text{IRDIS} & 5683 \pm 10 & 274.35 \pm 0.08 & 16.22 \pm 0.16 & 15.90 \pm 0.33 & 5661 \pm 7 & 274.87 \pm 0.08 & 16.14 \pm 0.13 & 15.54 \pm 0.19 \\
38 & \text{IRDIS} & 5744 \pm 6 & 243.75 \pm 0.05 & 14.49 \pm 0.07 & 14.33 \pm 0.10 & 5695 \pm 3 & 244.13 \pm 0.06 & 15.05 \pm 0.07 & 14.32 \pm 0.14 \\
39 & \text{IRDIS} & 5809 \pm 6 & 91.49 \pm 0.05 & 13.60 \pm 0.07 & 13.53 \pm 0.07 & 5839 \pm 2 & 90.94 \pm 0.05 & 13.57 \pm 0.06 & 13.47 \pm 0.05 \\
40 & \text{IRDIS} & 5882 \pm 11 & 285.40 \pm 0.10 & 16.08 \pm 0.20 & \text{\textemdash} & 5865 \pm 5 & 285.96 \pm 0.07 & 15.72 \pm 0.10 & 15.48 \pm 0.25 \\
41 & \text{IRDIS} & 5986 \pm 8 & 352.50 \pm 0.33 & 15.42 \pm 0.13 & 15.27 \pm 0.17 & 6032 \pm 5 & 352.85 \pm 0.31 & 15.59 \pm 0.10 & 15.13 \pm 0.13 \\
43 & \text{IRDIS} & 6226 \pm 9 & 312.07 \pm 0.09 & 15.72 \pm 0.17 & 15.56 \pm 0.21 & 6238 \pm 5 & 312.58 \pm 0.08 & 15.56 \pm 0.10 & 15.05 \pm 0.29 \\
44 & \text{IRDIS} & 6391 \pm 7 & 316.04 \pm 0.07 & 14.52 \pm 0.12 & 14.50 \pm 0.15 & 6414 \pm 4 & 316.61 \pm 0.07 & 14.56 \pm 0.09 & 14.22 \pm 0.13 \\
45 & \text{IRDIS} & 6456 \pm 8 & 351.89 \pm 0.30 & 15.47 \pm 0.14 & 15.13 \pm 0.16 & 6508 \pm 6 & 352.15 \pm 0.27 & 15.28 \pm 0.10 & 15.08 \pm 0.16 \\
46 & \text{IRDIS} & 6526 \pm 11 & 279.64 \pm 0.10 & 15.46 \pm 0.22 & \text{\textemdash} & 6511 \pm 6 & 280.16 \pm 0.08 & 15.57 \pm 0.13 & 14.94 \pm 0.27 \\
\hline        
\multicolumn{10}{c}{\text{Uncertain status}} \\
\hline        
7 & \text{IRDIS} & \text{\textemdash} & \text{\textemdash} & \text{\textemdash} & \text{\textemdash} & 2687 \pm 9 & 321.04 \pm 0.31 & 16.66 \pm 0.17 & 16.06 \pm 0.35 \\
8 & \text{IRDIS} & \text{\textemdash} & \text{\textemdash} & \text{\textemdash} & \text{\textemdash} & 2727 \pm 5 & 199.96 \pm 0.64 & 16.74 \pm 0.26 & 16.58 \pm 0.32 \\
10 & \text{IRDIS} & \text{\textemdash} & \text{\textemdash} & \text{\textemdash} & \text{\textemdash} & 3087 \pm 9 & 332.14 \pm 0.35 & 16.58 \pm 0.19 & 16.07 \pm 0.52 \\
11 & \text{IRDIS} & \text{\textemdash} & \text{\textemdash} & \text{\textemdash} & \text{\textemdash} & 3388 \pm 6 & 157.53 \pm 0.26 & 16.03 \pm 0.11 & 15.71 \pm 0.19 \\
20 & \text{IRDIS} & \text{\textemdash} & \text{\textemdash} & \text{\textemdash} & \text{\textemdash} & 4621 \pm 10 & 45.06 \pm 0.19 & 16.43 \pm 0.17 & 17.11 \pm 1.00 \\
25 & \text{IRDIS} & \text{\textemdash} & \text{\textemdash} & \text{\textemdash} & \text{\textemdash} & 5135 \pm 3 & 158.03 \pm 0.08 & 14.67 \pm 0.06 & 14.53 \pm 0.08 \\
29 & \text{IRDIS} & \text{\textemdash} & \text{\textemdash} & \text{\textemdash} & \text{\textemdash} & 5232 \pm 8 & 227.93 \pm 0.14 & 16.45 \pm 0.14 & \text{\textemdash} \\
30 & \text{IRDIS} & \text{\textemdash} & \text{\textemdash} & \text{\textemdash} & \text{\textemdash} & 5249 \pm 10 & 345.87 \pm 0.36 & 16.54 \pm 0.17 & \text{\textemdash} \\
34 & \text{IRDIS} & \text{\textemdash} & \text{\textemdash} & \text{\textemdash} & \text{\textemdash} & 5474 \pm 3 & 209.68 \pm 0.08 & 15.36 \pm 0.07 & 15.13 \pm 0.12 \\
35 & \text{IRDIS} & \text{\textemdash} & \text{\textemdash} & \text{\textemdash} & \text{\textemdash} & 5588 \pm 9 & 101.86 \pm 0.10 & 16.50 \pm 0.18 & 16.31 \pm 0.45 \\
42 & \text{IRDIS} & \text{\textemdash} & \text{\textemdash} & \text{\textemdash} & \text{\textemdash} & 6064 \pm 5 & 80.94 \pm 0.07 & 15.63 \pm 0.10 & 15.37 \pm 0.19 \\
\hline        
\end{array}
\]
}
\end{table*}

\end{appendix}

\listofobjects

\end{document}